\documentclass[prd,aps,reprint,preprintnumbers,nofootinbib,superscriptaddress,showkeys]{revtex4-1}
\usepackage{lipsum}
\usepackage{graphicx}
\usepackage{epsfig}
\usepackage{epstopdf}
\usepackage{hyperref}
\usepackage{amsmath}
\usepackage{mathrsfs}
\usepackage{amsfonts}
\usepackage{amssymb}
\usepackage{setspace}
\usepackage{verbatim}
\usepackage{color}
\usepackage{subfigure}

\begin{document}

\title{Contributions of box diagrams and $\Delta(1940)$ resonance to $K^0\Sigma^+$ photoproduction}

\author{Xuyang Liu}
\email{lxy\_gzu2005@126.com: corresponding author}
\affiliation{School of Mathematics and Physics, Bohai University, Liaoning, 121013, China}

\author{Daris Samart}
\email{darisa@kku.ac.th: corresponding author}
\affiliation{Department of Physics, Faculty of Science, Khon Kaen University,123 Mitraphap Rd., Khon Kaen, 40002, Thailand}
%\affiliation{Center of Excellence in High Energy Physics and Astrophysics, Suranaree University of Technology, Nakhon Ratchasima, 30000, Thailand}
%\affiliation{Kavli Institute for Theoretical Physics China at the Chinese Academy of Sciences, Beijing 100190, P. R. China}
%\affiliation{Institute of High Energy Physics, Chinese Academy of Sciences, Beijing 100049, P.R. China}

\author{Kai Xu}
\affiliation{School of Physics, Suranaree University of Technology, Nakhon Ratchasima, 30000, Thailand}
\affiliation{Center of Excellence in High Energy Physics and Astrophysics, Suranaree University of Technology, Nakhon Ratchasima, 30000, Thailand}

\author{Ayut Limphirat}
\affiliation{School of Physics, Suranaree University of Technology, Nakhon Ratchasima, 30000, Thailand}
\affiliation{Center of Excellence in High Energy Physics and Astrophysics, Suranaree University of Technology, Nakhon Ratchasima, 30000, Thailand}

\author{Yupeng Yan}
%\email{yupeng@sut.ac.th}
\affiliation{School of Physics, Suranaree University of Technology, Nakhon Ratchasima, 30000, Thailand}
\affiliation{Center of Excellence in High Energy Physics and Astrophysics, Suranaree University of Technology, Nakhon Ratchasima, 30000, Thailand}

\author{Qiang Zhao}
\affiliation{Institute of High Energy Physics, Chinese Academy of Sciences, Beijing 100049, P.R. China}
\affiliation{Theoretical Physics Center for Science Facilities, Chinese Academy of Sciences, Beijing 100049, P.R. China}
\date{\today}
\begin{abstract}
{We study first the box-diagram contribution to the $\gamma p\,\rightarrow \,K^0\Sigma^+$\, process to understand the anomaly of the kaon photoproduction cross section from CBELSA/TAPS experiment at Electron Stretcher Accelerator (ELSA), where the imaginary part of the scattering amplitude from the box-diagrams is  calculated by using Cutkosky's rules in the on-shell approximation while the real part of the amplitude is derived by dispersion relation calculations. Together with the results of the K-MAID model, the contribution of the box-diagrams fails to provide the sudden drop of the differential cross-section between the $K^*\,\Lambda$ and $K^*\,\Sigma$ thresholds. In addition, we include the $\Delta(1940)$ resonance in the process to complete the description of the differential cross-section. Combining the contributions from the K-MAID model, the box-diagrams and the $\Delta(1940)$ resonance, we have obtained the theoretical differential cross-section the $\gamma p\,\rightarrow \,K^0\Sigma^+$\, process, which is compatible with the CBELSA/TAPS experimental data.}
\end{abstract}

%%%%%%%%%%%%%%%%%%%%%%%%%%%%%%%%%%%%%%%
\maketitle
%\tableofcontents
\section{Introduction}
The study of photoproduction processes is an effective tool to investigate the internal structures of the baryon excitations \cite{Hey:1982aj,Glozman:1995da,Anisovich:2007bq,Klempt:2009pi,Aznauryan:2011qj,Crede:2013sze} as well as the hadronic interactions and its dynamical generated resonances and quasi-bound states \cite{Kaiser:1996js,Nacher:1998mi,Soyeur:2004ay,Liu:2008qx,Wang:2015jsa,Chen:2016qju,Guo:2017jvc,Liu:2019zoy}.
One of the theoretical milestone was accomplished by Kroll and Rundermann \cite{Kroll:1953vq}. The Kroll-Runderman theorem provides model-independent predictions of cross sections for pion photoproduction, i.e., $\gamma N \rightarrow \pi N$, in the threshold region by applying the gauge and Lorentz invariance. Later, the general formalism of scattering amplitudes is constructed by Chew, Goldberger, Low and Nambu \cite{Chew:1957tf}, known as CGLN amplitudes for single-pion photoproduction from the nucleon, which provides a systematic and convenient framework of theoretical calculations to compare with the experimental observables at the low-energies regime. %are in good agreement with experimental results at the low-energies regime.
Numerous research works on both theoretical and phenomenological approaches have been done so far, for example, phenomenological Lagrangian  \cite{Dalitz:1967fp,Thom:1966rm,Berends:1967vi,Walker:1968xu,Adelseck:1986fb,Workman:1989zw,Adelseck:1990ch,Drechsel:1992pn,Steininger:1996xw,Guidal:1997hy}, coupled-channel dynamics \cite{Kaiser:1996js,Borasoy:2007ku,JuliaDiaz:2007fa,Huang:2011as,Ronchen:2012eg} and quark models \cite{Copley:1969ft,Li:1995si,Li:1997gd,Zhao:2002id,Zhong:2011ti}.
%Since the properties of the excited states of the baryons reflect the dynamics and relevant degrees of freedom with them, it is clear that the baryon spectrum is considered to be a good place to study the strong interaction at low-energy where the QCD coupling constant is large in this region. Then, standard perturbative QFT methods is no longer applicable. It is well known that in the low-energy regime of quantum chromodynamics, as a result of confinement, the degrees of freedom are no longer quarks and gluons, but rather hadrons. A number of non-perturbative methods have been developed to work on this region. Among the various non-perturbative approaches, chiral perturbation theory (ChPT) \cite{Gasser:1984,Scherer:2012xha} is an effective field theory which is directly constructed from the chiral symmetry of QCD. In addition, the effective Lagrangian of the ChPT is a very powerful tool to study meson photoproduction.
One of the reasonably successful models of the meson photoproduction is the so-called MAID which is developed by a research group at Mainz University  \cite{Knochlein:1995qz,Drechsel:1998hk,Kamalov:2000en,Chiang:2001as,Chiang:2002vq,Drechsel:2007if,Hilt:2013fda} (see \cite{Tiator:2011pw,Tiator:2018pjq} for recent reviews) based on chiral effective Lagrangian in the Born approximation with observed resonant channels. The model has been extended to the strangeness sector of the meson photoproduction (called K-MAID \cite{Lee:1999kd}), and the theoretical predictions of the K-MAID are well consistent with the experimental data at the center of mass energy below 2 GeV \cite{Mart:1999ed,Bennhold:2000id,Bennhold:1999mt}.

The CBELSA/TAPS experiment at the Electron Stretcher Accelerator (ELSA) of Bonn University has investigated the excitation spectrum of the nucleon in the $\gamma p \rightarrow K^{0} \Sigma^{+}$ photoproduction reaction at low-energy region, i.e., the center of mass energy from 1.70 to 2.25 GeV \cite{Ewald:2011gw}. The differential cross section raises with increasing the energy up to the $K^{*}\Lambda$ threshold, and then drops rapidly in the energy region of 2.0074 $-$ 2.0855 GeV, and finally turns to a rather flat distribution at high energies. This result reveals the anomaly of the different cross-section at forward angle in the $K^*\Lambda$ and $K^*\Sigma$ regime. The K-MAID result is expected to describe the $\gamma p \rightarrow K^{0} \Sigma^{+}$ data. According to the differential cross-section in Ref. \cite{Ewald:2011gw}, however, one finds that the theoretical result of the K-MAID model is in line with the data at the very low energy region only. The modification of the parameters of the K-MAID (modified K-MAID) is also taken into account and the result of the modified K-MAID is compatible with the differential cross-section data at beyond the $K^*\Sigma$ threshold region. The discrepancy between the experimental data and the theoretical results from both the K-MAID and modified K-MAID models indicates that it is necessary to further improve the K-MAID model by adding new physics or more mechanism to the $\gamma p \rightarrow K^{0} \Sigma^{+}$ photoproduction reaction. The solution of this discrepancy has been speculated in \cite{Ewald:2011gw} that the contributions of the dynamically generated quasi-bound states for $K^*\Lambda$ and $K^*\Sigma$ should play crucial role to the differential cross-section in the $K^*\Lambda$ and $K^*\Sigma$ thresholds. To include this physical mechanism to the reaction, one may calculate the coupled-channel effects of the vector-meson and baryon interactions or box-diagram in the on-shell approximation of the intermediated states. In addition, the similar results of the $K^{0} \Sigma^{+}$ photoproduction have been found previously in Crystal Barrel experiment \cite{Castelijns:2007qt}.
%\begin{figure}
%\begin{center}
%\includegraphics[width=10cm,clip=true]{elsa.eps}%[width=6cm,height=5.5cm,angle=0]{cbelsa.eps}
%\end{center}
%\caption{ The differential cross-sections of the $\gamma p \rightarrow K^{0} \Sigma^{+}$ process is taken from Ref. \cite{Ewald:2011gw}. Solid and open squares are the data from the CBELSA/TAPS and Crystal Barrel \cite{Castelijns:2007qt} experiments respectively, and the dashed and solid lines are the theoretical results of the K-MAID model and the modified K-MAID model, respectively.}\label{CBELSA}
%\end{figure}

After the speculation of the anomaly of $K^0\Sigma^+$ photoproduction was proposed in Ref. \cite{Ewald:2011gw}, the cross-section of the $\gamma p \rightarrow K^{0} \Sigma^{+}$ reaction is investigated \cite{Ramos:2013wua},  by applying a chiral unitary coupled-channel approach with the effective chiral vector-meson and baryon interactions from hidden gauge formalism.
The results in Ref. \cite{Ramos:2013wua} exhibit the sudden drop of the cross-section in the region of the $K^*\Lambda$ and $K^*\Sigma$ thresholds and it is compatible with the experimental data.

In this work, we will carry out detailed calculations for the one-loop box-diagrams of the $\gamma p \rightarrow K^{0} \Sigma^{+}$ reaction in chiral effective Lagrangians by using the Cutkosky rules or the on-shell approximation of the intermediated states. In addition, we also include relevant Born diagrams with modified parameters of the K-MAID as well as contributions from $N^*$ and $\Delta^*$ resonances. We expect to shed some light on the understanding of the anomaly of the different cross-section from the CBELSA/TAPS experiment and the work would be of some value for the understanding of hadronic structures and interactions in the $K^0\Sigma$ photoproduction process.
We outline this work as follows: the ingredients of the box-diagrams and related formalisms are presented in the section II. The numerical results of the box-diagrams in on-shell approximation, modified K-MAID and additional baryon resonance are shown in the section III. In the last section, we will give the discussions and conclusions of this work.

\section{Formalism}
In this section, we present the formalisms for studying the differential cross-section in the $\gamma p\,\rightarrow\,K^0\Sigma^+$ process. We divide the ingredients of our formalisms into three parts as the box-diagrams, modified K-MAID and baryon resonance. The details of these ingredients are given in the following subsections.
\subsection{The box-diagram calculations in the on-shell scheme}
\subsubsection{Effective Lagrangian for kaon-photoproduction process}
\begin{figure}
\begin{center}
\includegraphics[width=9cm,clip=true]{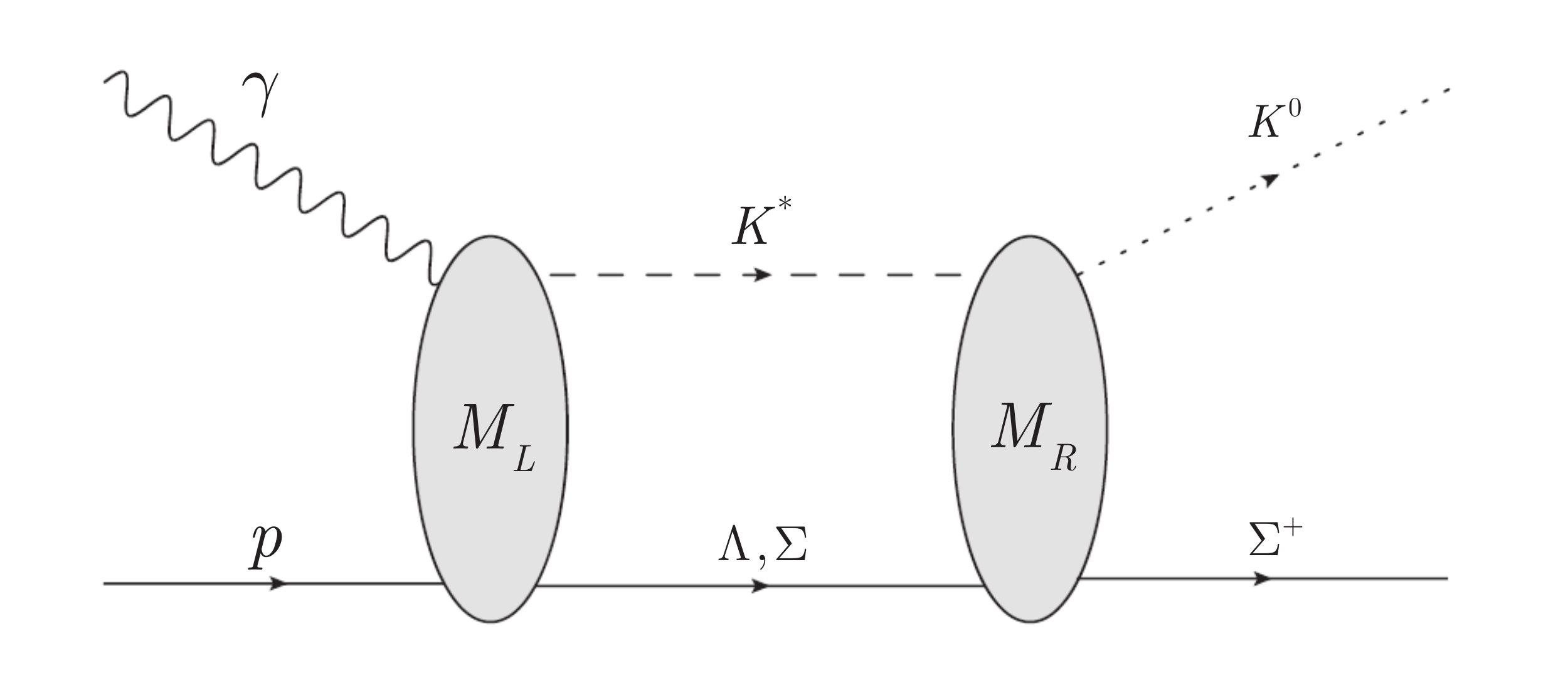}
\end{center}
\caption{Box-diagrams: Vector kaons and hyperons generated as the intermediated particles, and then inelastically re-scattered to the $K^0\Sigma^+$ final state.}
\label{box-diagram}
\end{figure}

The relevant Feynman box-diagrams for the $\gamma p\,\rightarrow\,K^0\Sigma^+$ process are depicted in Fig. \ref{box-diagram}. The chiral effective Lagrangian for the $\gamma\,K^*\,K$ coupling is given by \cite{Hyodo:2004vt,Oh:2003kw}
\begin{eqnarray}
\mathscr{L}_{\gamma\,\bar K^*\,K} &=& g_{\gamma\,\bar K^*\,K}\,\epsilon_{\mu\nu\alpha\beta}\,\partial^\mu\,A^\nu
\,\partial^\alpha\,\bar K^{*,\,\beta}\,K %+ \partial^\alpha\,\bar K^{*,\,\beta}\,K \Big)
+ {\rm h.c.}\,,
\end{eqnarray}
and the chiral effective Lagrangian for the $\pi\,K^*\,K$ coupling reads \cite{Hyodo:2004vt,Palomar:2002hk}
\begin{eqnarray}
\mathscr{L}_{\pi\,K^*\,\bar K} &=& -\,i\,g_{\pi\,K^*\,\bar K}\Big\{ \,\bar K\,\vec{\tau}\cdot\,\partial_\mu\,\vec\pi\,K^{*\,\mu}
\nonumber\\
&&\qquad\qquad\quad\; -\, (\partial_\mu\,\bar K)\,\vec{\tau}\cdot\,\vec\pi\,K^{*\,\mu}\, \Big\} + {\rm h.c.}\,.
\end{eqnarray}
For the meson-baryon interactions, we employ the axial-coupling vertices which are written in terms of the isospin basis \cite{Chiang:2004ye}
\allowdisplaybreaks
\begin{eqnarray}
\mathscr{L}_{K\,N\,\Lambda} &=& -\,\frac{g_{K\,N\,\Lambda}}{M_\pi} \,\bar \Lambda\,\gamma^\mu\,\gamma_5\,(\partial_\mu\,\bar K)\,N
+ {\rm h.c.} \,,
\nonumber\\
\mathscr{L}_{K\,N\,\Sigma} &=& -\,\frac{g_{K\,N\,\Sigma}}{M_\pi}\, \,\vec{\bar \Sigma}\,\cdot(\,\gamma^\mu\,\gamma_5\,\partial_\mu\,\bar K\,\vec\tau\,N)
+ {\rm h.c.} \,,
\nonumber\\
\mathscr{L}_{\pi\,\Lambda\,\Sigma} &=& -\,\frac{g_{\pi\,\Lambda\,\Sigma}}{M_\pi}\, \,(\vec{\bar \Sigma}\,\gamma^\mu\,\gamma_5\,\Lambda)\,\cdot\,\partial_\mu\,\vec{\pi} + {\rm h.c.} \,,
\nonumber\\
\mathscr{L}_{\pi\,\Sigma\,\Sigma} &=& i\,\frac{g_{\pi\,\Sigma\,\Sigma}}{M_\pi}\,\big(\vec{\bar \Sigma}\,\gamma^\mu\,\gamma_5\,\times\,\vec\Sigma \big)
\,\cdot\,\partial_\mu\,\vec\pi \,,
\end{eqnarray}
%One notes that $g_{MBB'}$ is axial-vector coupling which related to pseudoscalar coupling $G_{MBB'}$ via the relation $\frac{g_{MBB'}}{m_\pi} = \frac{G_{MBB'}}{M_B+M_{B'}}$ \cite{Chiang:2004ye}.
%In order to be consistence with the chiral Lagrangian of $\pi\,K^*\,K$\, couplings, one notes that the minus phase factor appears from the chiral perturbation theory \cite{Krause:xc,Scherer:2002tk}. In addition, coupling strengths of above Lagrangians are dimensionless parameters.
where the phase convention of mesons and baryons in the isospin basis is defined by
\begin{eqnarray}
N &=& \left(
\begin{array}{c}
 p \\
 n
\end{array}
\right)\,,~~~~
K =\left(
\begin{array}{c}
 K^+ \\
 K^0
\end{array}
\right)\,,~~~~
K^* =\left(
\begin{array}{c}
 K^{+\,*} \\
 K^{0\,*}
\end{array}
\right)\,,
\nonumber\\
\vec{\pi} &=& \left\{{\textstyle{\frac{1}{\sqrt{2}}}}\,(\pi^+ + \pi^-)\,,\;{\textstyle{\frac{i}{\sqrt{2}}}}\,(\pi^+ - \pi^-)\,,\;\pi^0 \right\}\,,
\nonumber\\
\vec{\Sigma}&=&\left\{{\textstyle{\frac{1}{\sqrt{2}}}}\,(\Sigma^+ + \Sigma^-)\,,\;{\textstyle{\frac{i}{\sqrt{2}}}}\,(\Sigma^+ - \Sigma^-)\,,\;\Sigma^0 \right\}\nonumber
\end{eqnarray}
The isospin generators $\vec\tau$\, are usual Pauli matrices,
\begin{equation}
\vec{\tau}=\left\{\tau_1\,,\;\tau_2\,,\;\tau_3 \right\}
=\left\{
\left(
\begin{array}{cc}
 0 & 1 \\
 1 & 0
\end{array}
\right)\,,\;
\left(
\begin{array}{cc}
 0 & -i \\
 i & 0
\end{array}
\right)\,,\;
\left(
\begin{array}{cc}
 1 & 0 \\
 0 & -1
\end{array}
\right)\right\}.\nonumber
\end{equation}
\subsubsection{Imaginary part of the scattering amplitudes in the on-shell scheme}
We calculate here the scattering amplitudes of the box-diagrams for $K^0\Sigma$-photoproduction in the on-shell approximation.
First of all, we assign the four-momenta for each particles as
\begin{eqnarray}
&&\gamma = \gamma(k)\,,\quad p = p(p)\,,\quad K^0 = K^0(q)\,,\quad
\nonumber\\
&&\Sigma^+ = \Sigma(\bar p)\,,
\quad K^* = K^*(l)\,.
\end{eqnarray}
In this work, we apply the Cutkosky rules to our box-diagram calculations. The formalism is to calculate the imaginary part of scattering amplitudes based on the optical theorem of the Feynman diagrams (for more detailed discussion see Ref. \cite{Peskin:1995ev}). The physical meaning of the scheme is that there are re-scattering effects between mesons and baryons in the intermediate states. The imaginary part of scattering amplitudes of the box-diagrams takes the generic form
\begin{widetext}
\begin{eqnarray}
{\rm Im}\,\mathcal{M}_{\rm box} = -\frac{\widetilde{l}}{32\,\pi^2\,\sqrt{s}}\int d\,\Omega\,\sum_{K^{*},Y}\,M_L\big(\gamma p\,\rightarrow\,K^*Y\big)\,M_R^\dagger\big(K^0\Sigma^+\rightarrow K^*Y\,\big)
\end{eqnarray}
\end{widetext}
where the summation of overall intermediated particles spin is implied, $Y=\Lambda,\,\Sigma$, and $\widetilde{l}$ is the 3-momentum magnitude in the c.m. frame of intermediate $K^*$ particle, reading as
\begin{eqnarray}
\widetilde{l}=\sqrt{\frac{\big[s - (m_{K^{*}} - m_Y)^2 \big] \big[s - (m_{K^{*}} + m_Y)^2 \big]}{4\,s}}\,.
\end{eqnarray}
We obtain, from the above effective Lagrangians, the amplitudes $M_L$ and $M_R$\, (see Fig. 2 for corresponding Feynman diagrams of $M_L$\, and $M_R$),
\begin{widetext}
\begin{eqnarray}
M_L^{(1)}\big(\gamma p\,\rightarrow\,K^{*,+} \Lambda\big) &=& -\,g_{\gamma\,\bar K^{*}\,K}^{(c)}\,\frac{g_{K\,N\,\Lambda}}{m_\pi}\,\epsilon_{\mu\nu\alpha\beta}\,k^\mu\,\epsilon_{(\gamma)}^\nu\,
l^\alpha\,\epsilon_{(K^*)}^{*\beta}\,\frac{\bar u_\Lambda\,(k\!\!\!/ - l\!\!\!/)\,\gamma_5\,u_p}{(k-l)^2 - m_{\bar K}^2}\,F_{K}^2(k-l)\,,
\label{ML1}\\
M_L^{(2)}\big(\gamma p\,\rightarrow\,K^{*,0} \Sigma^+\big) &=& -\sqrt{2}\,g_{\gamma\,\bar K^{*}\,K}^{(0)}\,\frac{g_{K\,N\,\Sigma}}{m_\pi}\,\epsilon_{\mu\nu\alpha\beta}\,k^\mu\,\epsilon_{(\gamma)}^\nu\,
l^\alpha\,\epsilon_{(K^*)}^{*\beta}\,\frac{\bar u_\Sigma\,(k\!\!\!/ - l\!\!\!/)\,\gamma_5\,u_p}{(k-l)^2 - m_{\bar K}^2}\,F_{K}^2(k-l)\,,
\label{ML2}\\
M_L^{(3)}\big(\gamma p\,\rightarrow\,K^{*,+} \Sigma^0\big) &=& -\,g_{\gamma\,\bar K^{*}\,K}^{(c)}\,\frac{g_{K\,N\,\Sigma}}{m_\pi}\,\epsilon_{\mu\nu\alpha\beta}\,k^\mu\,\epsilon_{(\gamma)}^\nu\,
l^\alpha\,\epsilon_{(K^*)}^{*\beta}\,\frac{\bar u_\Sigma\,(k\!\!\!/ - l\!\!\!/)\,\gamma_5\,u_p}{(k-l)^2 - m_{\bar K}^2}\,F_{K}^2(k-l)\,,
\label{ML3}\\
M_R^{(1),\dagger}\big(K^0 \Sigma^+\rightarrow\,K^{*,+} \Lambda\,\big) &=& 2\,\sqrt{2}\,g_{\pi\,K^{*}\,\bar K}\,\frac{g_{\pi\,\Lambda\,\Sigma}}{m_\pi}\,q_\mu\,\epsilon_{(K^*)}^{\mu}\,\frac{\bar u_\Sigma\,(l\!\!\!/ - q\!\!\!/)\,\gamma_5\,u_\Lambda}{(l - q)^2 -m_\pi^2}\,F_\pi^2(l - q)\,,
\label{MR1}\\
M_R^{(2),\dagger}\big(K^0 \Sigma^+\rightarrow\,K^{*,0} \Sigma^+\big) &=& -\,2\,g_{\pi\,K^{*}\,\bar K}\,\frac{g_{\pi\,\Sigma\,\Sigma}}{m_\pi}\,q_\mu\,\epsilon_{(K^*)}^{\mu}\,\frac{\bar u_\Sigma\,(l\!\!\!/ - q\!\!\!/)\,\gamma_5\,u_\Sigma}{(l - q)^2 -m_\pi^2}\,F_\pi^2(l - q)\,,
\label{MR2}\\
M_R^{(3),\dagger}\big(K^0 \Sigma^+\rightarrow\,K^{*,+} \Sigma^0\big) &=& 2\,g_{\pi\,K^{*}\,\bar K}\,\frac{g_{\pi\,\Sigma\,\Sigma}}{m_\pi}\,q_\mu\,\epsilon_{(K^*)}^{\mu}\,\frac{\bar u_\Sigma\,(l\!\!\!/ - q\!\!\!/)\,\gamma_5\,u_\Sigma}{(l - q)^2 -m_\pi^2}\,F_\pi^2(l - q)\,,
\label{MR3}
\end{eqnarray}
\end{widetext}
where $u_{p,\Sigma,\Lambda}$ stand for the spinors of the proton, $\Sigma$ and $\Lambda$ baryons respectively, and $\epsilon_{(\gamma,K^*)}^{\mu}$ are the polarization vectors of the photon and vector $K^*$ meson respectively. $F_X(q)$ are form-factors introduced to regularize the high momentum behavior of the meson-baryon couplings, taking the form,
\begin{eqnarray}
F_X(q) = \frac{\Lambda_X^2 - m_X^2}{\Lambda_X^2 - q^2}\,,
\end{eqnarray}
where $X = K\,,~\pi$\,. Here we take $\Lambda_{K} = 900$ MeV and $\Lambda_{\pi} = 450$ MeV. Note that the identity $l^\mu\,\epsilon_{\mu}^*(l\,,m_{K^*}) = 0$\, is applied to above equations.

As the data analysis of the experimental results and the speculation from Ref. \cite{Ewald:2011gw} indicate that the $t$-channel is dominant, we consider the $t$-channel amplitudes for the on-shell intermediate states only.
Then the imaginary part of scattering amplitude can be written in the following form,
\begin{eqnarray}
{\rm Im}\,\mathcal{M}_{\rm box} = {\rm Im}\,\mathcal{M}_{\rm box}^{(1)} + {\rm Im}\,\mathcal{M}_{\rm box}^{(2)} + {\rm Im}\,\mathcal{M}_{\rm box}^{(3)}\,.
\end{eqnarray}
The ${\rm Im}\,\mathcal{M}_{\rm box}^{(1)}$ amplitude reads,
\allowdisplaybreaks
\begin{widetext}
\begin{eqnarray}\label{box1}
{\rm Im}\,\mathcal{M}_{\rm box}^{(1)} &=& -\frac{\widetilde{l}}{32\,\pi^2\,\sqrt{s}}\int d\,\Omega\,\sum_{K^{*,+},\Lambda}\,M_L^{(1)}\big(\gamma p \rightarrow K^{*,+}\Lambda\big)\,M_R^{(1)\;\dagger}\big(K^0\Sigma^+\rightarrow K^{*,+}\Lambda\big)
\nonumber\\
&=& \frac{\widetilde{l}}{64\,\pi^2\,\sqrt{s}}\,G_1 \int d\,\Omega\;\frac{F_{K}^2(k-l)}{(k-l)^2 - m_{\bar K}^2}
\,\frac{F_\pi^2(l - q) }{(l - q)^2 -m_\pi^2}
\\
&\;&\times\,\bar u_\Sigma\,
\big(k\!\!\!/\,\epsilon\!\!\!/\,q\!\!\!/\,l\!\!\!/ - l\!\!\!/\,k\!\!\!/\,\epsilon\!\!\!/\,q\!\!\!/- k\cdot q\,l\!\!\!/\,\epsilon\!\!\!/ + k\cdot l\,\epsilon\!\!\!/\,l\!\!\!/ - q\cdot\epsilon\,k\!\!\!/\,l\!\!\!/ + q\cdot\epsilon\,l\!\!\!/\,k\!\!\!/ \big)\,\gamma_5
\,(l\!\!\!/ - q\!\!\!/)\,\gamma_5\,(p\!\!\!/ + k\!\!\!/ - l\!\!\!/ + m_\Lambda)\,(k\!\!\!/ - l\!\!\!/)\,\gamma_5\,u_p\,, \nonumber
\end{eqnarray}
where \,$l^2 = m_{K^*}^2$,\, $k^2 = 0$,\, and \,$G_1=-\,2\,\sqrt{2}\,g_{\gamma\,\bar K^{*}\,K}^{(c)}\,g_{\pi\,K^{*}\,\bar K}\,\frac{g_{K\,N\,\Lambda}}{m_\pi}\,\frac{g_{\pi\,\Lambda\,\Sigma}}{m_\pi}\,$, and the scalar dot production of the four-vector is used by the notation $A\cdot B = A_\mu\,B^\mu  = g_{\mu\nu}A^\mu\,B^\nu = g^{\mu\nu}A_\mu\,B_\nu$\,. With the same manner, one obtains the imaginary part of scattering amplitudes, ${\rm Im}\,\mathcal{M}_{\rm box}^{(2)}$\, and ${\rm Im}\,\mathcal{M}_{\rm box}^{(3)}$. They are given by
\begin{eqnarray}\label{box2}
{\rm Im}\,\mathcal{M}_{\rm box}^{(2)} &=& -\frac{\widetilde{l}}{32\,\pi^2\,\sqrt{s}}\int d\,\Omega\,\sum_{K^{*,0},\Sigma^+}\,M_L^{(2)}\big(\gamma p\,\rightarrow\,K^{*,0} \Sigma^+\big)\,M_R^{(2)\;\dagger}\big(K^0 \Sigma^+\,\rightarrow\,K^{*,0} \Sigma^+\big)
\nonumber\\
&=& \frac{\widetilde{l}}{64\,\pi^2\,\sqrt{s}}\,G_2 \int d\,\Omega\;\frac{F_{K}^2(k-l)}{(k-l)^2 - m_{\bar K}^2}
\,\frac{F_\pi^2(l - q) }{(l - q)^2 -m_\pi^2}
\\
&\;&\times\,\bar u_\Sigma\,
\big(k\!\!\!/\,\epsilon\!\!\!/\,q\!\!\!/\,l\!\!\!/ - l\!\!\!/\,k\!\!\!/\,\epsilon\!\!\!/\,q\!\!\!/- k\cdot q\,l\!\!\!/\,\epsilon\!\!\!/ + k\cdot l\,\epsilon\!\!\!/\,l\!\!\!/ - q\cdot\epsilon\,k\!\!\!/\,l\!\!\!/ + q\cdot\epsilon\,l\!\!\!/\,k\!\!\!/ \big)\,\gamma_5
\,(l\!\!\!/ - q\!\!\!/)\,\gamma_5\,(p\!\!\!/ + k\!\!\!/ - l\!\!\!/ + m_\Sigma)\,(k\!\!\!/ - l\!\!\!/)\,\gamma_5\,u_p \,, \nonumber
\end{eqnarray}
where \,$G_2= 2\,\sqrt{2}\,g_{\gamma\,\bar K^{*}\,K}^{(0)}\,g_{\pi\,K^{*}\,\bar K}\,\frac{g_{K\,N\,\Sigma}}{m_\pi}\,\frac{g_{\pi\,\Sigma\,\Sigma}}{m_\pi}\,$,
and
\begin{eqnarray}\label{box3}
{\rm Im}\,\mathcal{M}_{\rm box}^{(3)} &=& -\frac{\widetilde{l}}{32\,\pi^2\,\sqrt{s}}\int d\,\Omega\,\sum_{K^{*,+},\Sigma^0}\,M_L^{(3)}\big(\gamma p\,\rightarrow\,K^{*,+} \Sigma^0\big)\,M_R^{(3)\;\dagger}\big(K^0 \Sigma^+\,\rightarrow\,K^{*,+} \Sigma^0 \big)
\nonumber\\
&=& \frac{\widetilde{l}}{64\,\pi^2\,\sqrt{s}}\,G_3 \int d\,\Omega\;\frac{F_{K}^2(k-l)}{(k-l)^2 - m_{\bar K}^2}
\,\frac{F_\pi^2(l - q) }{(l - q)^2 -m_\pi^2}
\\
&\;&\times\,\bar u_\Sigma\,
\big(k\!\!\!/\,\epsilon\!\!\!/\,q\!\!\!/\,l\!\!\!/ - l\!\!\!/\,k\!\!\!/\,\epsilon\!\!\!/\,q\!\!\!/- k\cdot q\,l\!\!\!/\,\epsilon\!\!\!/ + k\cdot l\,\epsilon\!\!\!/\,l\!\!\!/ - q\cdot\epsilon\,k\!\!\!/\,l\!\!\!/ + q\cdot\epsilon\,l\!\!\!/\,k\!\!\!/ \big)\,\gamma_5
\,(l\!\!\!/ - q\!\!\!/)\,\gamma_5\,(p\!\!\!/ + k\!\!\!/ - l\!\!\!/ + m_\Sigma)\,(k\!\!\!/ - l\!\!\!/)\,\gamma_5\,u_p \,, \nonumber
\end{eqnarray}
where \,$G_3= -\,2\,g_{\gamma\,\bar K^{*}\,K}^{(c)}\,g_{\pi\,K^{*}\,\bar K}\,\frac{g_{K\,N\,\Sigma}}{m_\pi}\,\frac{g_{\pi\,\Sigma\,\Sigma}}{m_\pi}\,$.
\begin{figure}
\begin{center}\label{MLR}$
\begin{array}{c}
\includegraphics[width=11cm,clip=true]{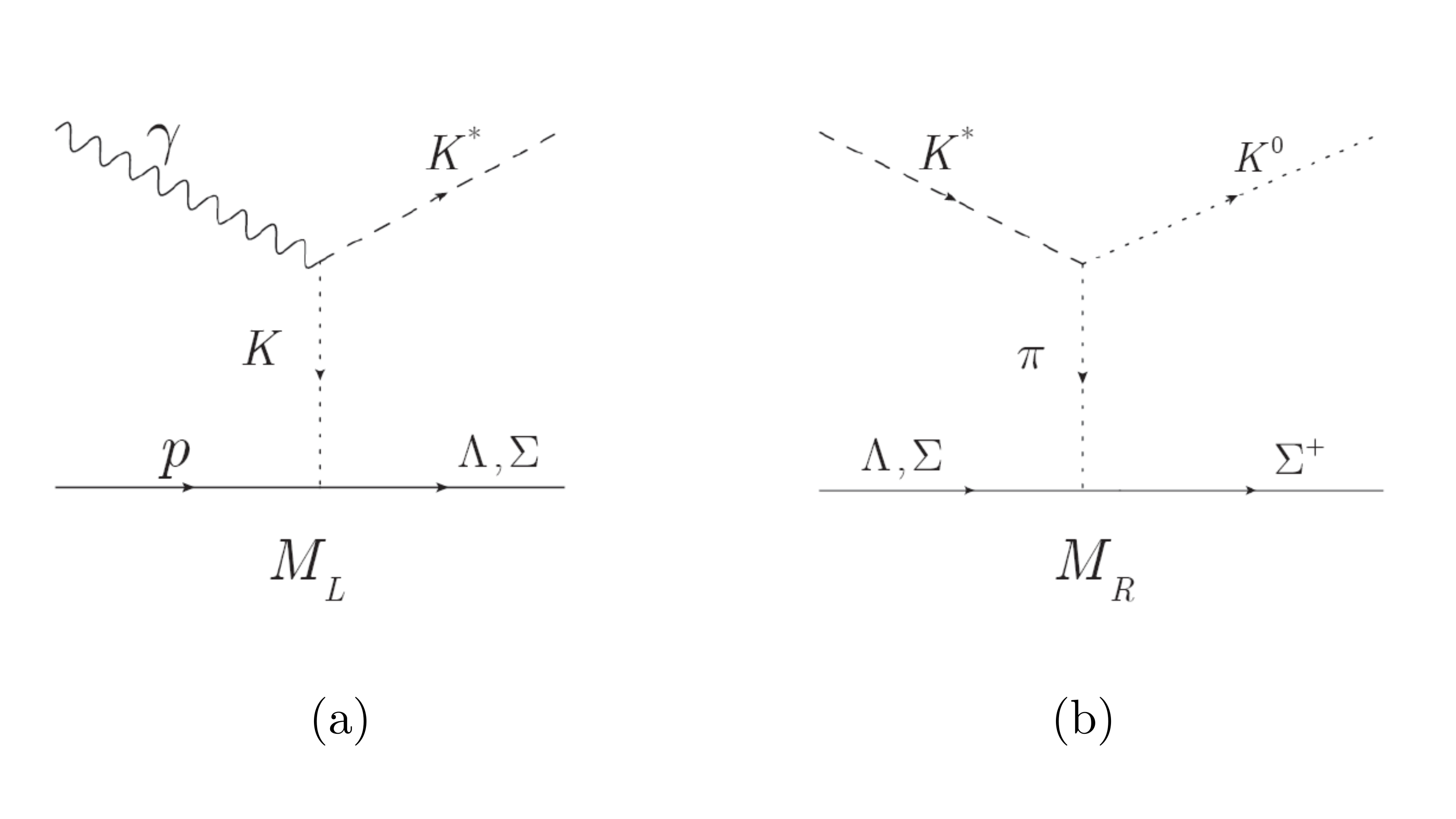}
\end{array}$
\end{center}
\caption{Feynman diagrams with all possible $K^*Y$ intermediate states in $t$-channel: (a) for $M_L^{(1,2,3)}\big(\gamma p\,\rightarrow\,K^*Y\big)$ and (b) for $M_R^{(1,2,3)}\big( K^*Y\rightarrow K^0\Sigma^+\big)$.}
\end{figure}
\end{widetext}

It is convenient to write the scattering amplitudes in terms of  the CGLN amplitudes which have been widely employed to study the photoproduction processes \cite{Chew:1957tf}. The transition amplitudes in Eqs. (\ref{box1},\ref{box2},\ref{box3}) can be expressed in terms of the CGLN amplitudes,
\begin{eqnarray}
&&{\rm Im}\,\mathcal{M}_{\rm box}^{(i)}= \int_{-1}^1 d(\cos\theta_{\rm in})\int_0^{2\pi} d\phi_{\rm in}
\nonumber\\
&&\qquad\times\,\Big(\mathcal{F}_1^{{\rm box}\,(i)} \vec\sigma\cdot\vec\varepsilon + \mathcal{F}_2^{{\rm box}\,(i)} i\,(\vec\sigma\cdot\hat q\,)\,(\vec\varepsilon\times\hat k\,)\cdot\,\vec\sigma
\nonumber\\
&&\qquad +\, \mathcal{F}_3^{{\rm box}\,(i)}(\vec\sigma\cdot\hat k\,)\,(\hat q\cdot\vec\varepsilon\,) + \mathcal{F}_4^{{\rm box}\,(i)}(\vec\sigma\cdot\hat q\,)\,(\hat q\cdot\vec\varepsilon\,)\Big)\,,
\end{eqnarray}
where index $i = 1,2,3$\,. The detailed derivation of the CGLN amplitudes of the box-diagrams is shown in the appendix, and the generic form of the CGLN amplitudes ($\mathcal{F}_i$) is given in Eq. (\ref{CGLN}). In the above equation, we have defined the kinematic variables in the c.m. frame. The (real) photon polarization has the component as
\,$\epsilon^\mu = \big( \,0\,,~\vec\varepsilon\;\big)$, and the components of the 3-momenta $\vec k\,,~\vec{l}$ and $\vec q$\, are given by
\begin{eqnarray}
\vec k &=& \widetilde{k}\,\big(\,0\,,~0\,,~1 \,\big)\,,
\nonumber\\
\vec{l} &=& \widetilde{l}\,\big(\,\sin\theta_{\rm in}\,\cos\phi_{\rm in}\,,~\sin\theta_{\rm in}\,\sin\phi_{\rm in}\,,~\cos\theta_{\rm in}\,\big)\,,
\nonumber\\
\vec q &=& \widetilde{q}\,\big(\,\sin\theta\,,~0\,,~\cos\theta\,\big)\,,
\end{eqnarray}
here we used notation $|\vec k\;|=\widetilde{k}$\,, $|\vec{l}\;|=\widetilde{l}$\, and $|\vec{q}\;|=\widetilde{q}$\, and all kinematic variables are in c.m. frame i.e. $\vec k = -\vec p$\, and $\vec q = -\vec{\bar p}$\,. In addition, the three-momenta, $\widetilde{k}$\,, and $\widetilde{q}$\, are written by
\begin{eqnarray}
\widetilde{k} &=& \widetilde{p} = \frac{\sqrt{\big[s - m_p^2 \big] \big[s - m_p^2 \big]}}{2\sqrt{s}} \,,
\nonumber\\
\widetilde{q} &=& \widetilde{\bar p} = \frac{\sqrt{\big[s - (m_{\Sigma} - m_{K})^2 \big] \big[s - (m_{\Sigma} + m_{K})^2\big]}}{2\sqrt{s}}\,.
\end{eqnarray}
In this work, the input parameters of the on-shell box-diagram calculation are taken from literatures,
$g_{\pi K^* \bar K} = -\,3.26$ \cite{Usov:2005wy}\,, $g_{\gamma K^* \bar K}^{(c)} = 0.254 ~{\rm GeV}^{-1}$ \cite{Oh:2003kw}\,,
$g_{\gamma K^* \bar K}^{(0)} = 0.388~ {\rm GeV}^{-1}$ \cite{Oh:2003kw}\,, $g_{K N\Lambda} = -\,0.95$ \cite{Chiang:2004ye}\,,
$g_{K N\Sigma} = 0.27$ \cite{Chiang:2004ye}\,, $g_{\pi\Lambda\Sigma} = 0.71$ \cite{Chiang:2004ye}\,, $g_{\pi\Sigma\Sigma} = 0.74$\cite{Chiang:2004ye}\,, $\Lambda_{K} = 900$ {\rm MeV}\,, $\Lambda_{\pi} = 450$ {\rm MeV}\,. The imaginary part of the box-diagram amplitudes will be calculated numerically and also employed to obtain the real part by using the dispersion relation in the next subsection.

\subsubsection{The real part of the box diagram amplitudes}
The intermediate states would contribute the dispersive (real) part of the box-diagram (loop) amplitudes beyond the $K^*\Lambda$ and $K^*\Sigma$ thresholds. The real part can be determined via  the dispersion relation \cite{Guo:2012tj},
\begin{widetext}
\begin{eqnarray}
{\rm Re}\,\mathcal{M}_{\rm box} = \frac{1}{2\,\pi\,i}\,\mathcal{P}\left( \int_{M_{K^*\Lambda}^2}^{M_{K^*\Sigma}^2}\,\frac{{\rm Im}\,\mathcal{M}_{\rm box}(s')}{s-s'}\,d\,s' + \int_{M_{K^*\Sigma}^2}^{\infty }\,\frac{{\rm Im}\,\mathcal{M}_{\rm box}(s')}{s-s'}\,d\,s'\right),
\label{Re-M}
\end{eqnarray} 
\end{widetext}%({\rm th}_{K^*\Sigma} + 150 {\rm MeV})^2
where $\mathcal{P}$ stands for the principal value of the complex integration, and $M_{K^*\Lambda}$ and $M_{K^*\Sigma}$ are the thresholds of $K^*\Lambda$ and $K^*\Sigma$ masses respectively. The real part amplitudes are calculated numerically and we will show the results in the next section.
%%%%%%%%%%%%%%%%%%%%%%%%%%%%%%%%%%%%%%%%%%%%%%%%%%%%%%%%%%%%%%%%%%%%%%%%%%%%%%%%%%%%%%%%%%%%%%%%%%%%%%%%%%%%%%%%%%%%%%%%%%%%%%%%%%%%%%
%\begin{figure}[t]
%        \includegraphics[width=10cm,clip=true]{modKMAID.eps}
%        \caption{The numerical result of differential cross section from modified K-MAID with $G1_{\rm S_{31}}(1900) = 0.8\,,~ G1_{\rm P_{31}}(1910) = 0.6\,,~ G2_{\rm S_{31}}(1900) = 0.8\,,~ G2_{\rm P_{31}}(1910) = 0.6$\,.} \label{kmaid}
%\end{figure}
%%%%%%%%%%%%%%%%%%%%%%%%%%%%%%%%%%%%%%%%%%%%%%%%%%%%%%%%%%%%%%%%%%%%%%%%%%%%%%%%%%%%%%%%%%%%%%%%%%%%%%%%%%%%%%%%%%%%%%%%%%%%%%%%%%%%%
\subsection{Modified K-MAID}
In this work, we will use the K-MAID model as the first-order approximation to the $K^0\Sigma^+$ photoproduction (see \cite{Tiator:2018pjq} for the latest status report). The K-MAID contains all possible Born amplitudes (tree-level) with $S$, $P$ and $D$ wave excitation states $N^*$ up to 1910 MeV. It has been demonstrated in Ref. \cite{Ewald:2011gw} that the K-MAID can not satisfactorily describe the data. However, the K-MAID is still a  good model and useful for studying the $\gamma p\rightarrow K^0\Sigma^+$ reaction below 1900 MeV by adjusting the model parameters. Moreover, the K-MAID model provides the CGLN amplitudes which can be incorporated to our box-diagram results directly. We extract the numerical value of CGLN amplitudes from the website \texttt{www.kph.uni-mainz.de/MAID/kaon/} with modified parameters in the K-MAID model,
\begin{eqnarray}
&& G1_{\rm S_{31}}(1900) = 0.8\,,\qquad G1_{\rm P_{31}}(1910) = 0.6\,,
\nonumber\\
&& G2_{\rm S_{31}}(1900) = 0.8\,,\qquad G2_{\rm P_{31}}(1910) = 0.6\,.
\label{kmaid-parameters}
\end{eqnarray}
The result of the modified K-MAID with respect to the data is depicted as dashed line in Fig. \ref{diff-cross-plots} (upper panel). With the proper modification of the K-MAID parameters in Eq. (\ref{kmaid-parameters}), the result is compatible with the data up to c.m. frame energy around 1920 MeV. However, the modified K-MAID still mismatches the experimental results beyond 1920 MeV. This requires additional mechanism such as box-diagrams and baryon resonances to understand the data.
%%%%%%%%%%%%%%%%%%%%%%%%%%%%%%%%%%%%%%%%%%%%%%%%%%%%%%%%%%%%%%%%%%%%%%%%%%%%%%%%%%%%%%%%%%%%%%%%%%%%%%%%%%%%%%%%%%%%%%%%%%%%%%%%%%%%%
\subsection{$\Delta(1940)$ resonance}
%According to the anomaly of the differential cross-section of the $K^0\Sigma^+$ photoproduction in Ref. \cite{Ewald:2011gw}, we found that the modified K-MAID results do not provide the peak of the differential cross-section below the $K^*\Lambda$ threshold around 1960-1980 MeV and the box-diagram amplitudes in the on-shell approximation also start contributions at the $K^*\Lambda$ threshold.
To improve the K-MAID model results of the differential cross section below the $K^*\Lambda$ threshold around 1960-1980 MeV, we propose to include the $\Delta(1940)$ resonance which contributes via the process $\gamma p\;\rightarrow\;\Delta(1940)\;\rightarrow\,K^0\Sigma^+$.  Here we take the vertices and propagators from Refs. \cite{Adelseck:1990ch,David:1995pi,Mart:2015jof}. In particular, we follow the notation and convention for analytical expressions from Ref. \cite{Mart:2015jof} which are compatible with the results from K-MAID model. For more detailed formalism, we refer to Refs. \cite{Adelseck:1990ch,David:1995pi,Mart:2015jof}. The coupling constants of the corresponding effective Lagrangians of the $\Delta(1940)$ resonance in the $K^0\Sigma$ photoproduction will be estimated in the subsequent subsections.
\subsubsection{Electromagnetic couplings}
In this subsection, we will fix the electromagnetic couplings $\kappa_1$ and $\kappa_2$ that appeared in the effective Lagrangians \cite{Adelseck:1990ch} and \cite{David:1995pi} (Lyon and Saclay groups). Since the coupling of a spin-1 photon and a spin-1/2 nucleon to a spin-3/2
resonance can be constructed in two different ways, there are two helicity amplitudes,
\begin{eqnarray}
A_{1/2} &=& \frac{e}{4\,m_R}\,\sqrt{\frac{m_R^2 - m_p^2}{3\,m_p}}\left( \kappa_1 + \frac{m_R}{4\,m_p^2}\,(m_R - m_p)\,\kappa_2\right) ,
\nonumber\\
A_{3/2} &=& \frac{e}{4\,m_p}\,\sqrt{\frac{m_R^2 - m_p^2}{m_p}}\left( \kappa_1 - \frac{1}{4\,m_p}\,(m_R - m_p)\,\kappa_2\right) .
\nonumber\\
\end{eqnarray}
The numerical values of $A_{1/2}$ and $A_{3/2}$ are taken from PDG for the $\Delta$(1940) electromagnetic decay. To fix the electromagnetic coupling constants, one can convert the above equations into an appropriate form as
\begin{eqnarray}
e\,\kappa_1 &=& \frac{4\,m_p\,m_R}{m_R + m_p}\sqrt{\frac{m_p}{m_R^2 - m_p^2}}\left( \sqrt{3}\,A_{1/2} + A_{3/2}\right) ,
\nonumber\\
e\,\kappa_2 &=& \frac{16\,m_p^2\,m_R}{m_R^2 - m_p^2}\sqrt{\frac{m_p}{m_R^2 - m_p^2}}\left( \sqrt{3}\,A_{1/2} - \frac{m_p}{m_R}\, A_{3/2}\right) .
\nonumber\\
\label{Delta1940-EM-couplings}
\end{eqnarray}
Taking as inputs the known parameters from PDG \cite{Tanabashi:2018oca},
\begin{eqnarray}
m_R &=& 2.000 \quad {\rm GeV}\,,
\quad
m_p = 0.983 \quad {\rm GeV}\,,
\nonumber\\
A_{1/2} &=& 0.170_{-0.080}^{+0.110} \quad{\rm GeV}^{-1/2}\,,
\nonumber\\
A_{3/2} &=& 0.150 \pm 0.080 \quad{\rm GeV}^{-1/2}\,.
\label{Delta-1940-helicity-amp}
\end{eqnarray}
By using Eqs. (\ref{Delta1940-EM-couplings}) and (\ref{Delta-1940-helicity-amp}), we estimate the electromagnetic coupling constants as
\begin{eqnarray}
e\,\kappa_1 &=& 0.667_{-0.328}^{+0.406} \,,
\nonumber\\
e\,\kappa_2 &=& 1.281_{-0.576}^{+0.877} \,.
%e\,\kappa_1 &=& -\,0.31 \,,
%\nonumber\\
%e\,\kappa_2 &=& -\,0.82 \,.
\end{eqnarray}
To make our prediction compatible with the CBELSA/TAPS data, we will use the values $e\,\kappa_1 = 0.34 $ and $e\,\kappa_2 = 0.82 $ in the latter calculation.

\subsubsection{Hadronic decay}
With the help of the effective Lagrangian we introduced in the first section, the amplitude of the $\Delta({\rm 1940})\,\rightarrow\,K\Sigma$ decay process is derived as
\begin{eqnarray}
\mathcal{M}_{\Delta(1940)\,\rightarrow\,K\Sigma} = i\,\frac{f_{KYR}}{m_K}\,\bar u(m_Y)\,p_{K}^\mu\,\gamma_5\,u_\mu(m_R)\,.
\end{eqnarray}
The decay width in the $\Delta(1940)$ rest frame is written as
\begin{eqnarray}
\Gamma_{\Delta(1940)\,\rightarrow\,K\Sigma} = \left( \frac{f_{KYR}}{m_K}\right)^2 \frac{|\vec p_K\,|^3}{12\,\pi}\left(\frac{E_Y + m_Y}{m_Y}\right) ,
\end{eqnarray}
where the magnitude of the 3-momentum $\vec p_K$ is defined as
\begin{eqnarray}
|\vec p_K\,| = \frac{\sqrt{\left[ m_R^2 - (m_Y + m_K)^2 \right] \left[ m_R^2 - (m_Y - m_K)^2 \right]}}{2\,m_R}\,.
\nonumber\\
\end{eqnarray}
The strong decay width, $\Gamma_{\Delta({\rm 1940})\,\rightarrow\,K\Sigma}$
is estimated, based on the PDG parameters in \cite{Agashe:2014kda} and the data in Ref. \cite{Candlin:1983cw},
\begin{eqnarray}
&&\left(\Gamma_{\Delta(1940)\,\rightarrow\,\pi\,N}\, \Gamma_{\Delta(1940)\,\rightarrow\,K\,\Sigma}\right)^{1/2}/\,\Gamma_{\rm total} < 0.015\,,
\nonumber \\
&&\;\;\Gamma_{\rm total}= 450 \pm 100 ~{\rm GeV}\,,
\nonumber\\
&&\;\;\Gamma_{\Delta(1940)\,\rightarrow\,\pi\,N}/\,\Gamma_{\rm total} = 5 \pm 2 ~ \%\,.
\end{eqnarray}
One finds
\begin{eqnarray}
\Gamma_{\Delta(1940)\rightarrow K\Sigma} < 3.4 \quad {\rm MeV}\,,
\end{eqnarray}
where we have used the central value for $\Gamma_{\rm total}$ but $\Gamma_{\Delta(1940)\,\rightarrow\,\pi\,N}/\,\Gamma_{\rm total} = 3 ~ \%$.
Then we fix the value of coupling constant $f_{KYR}$ with the relative sign, $\varepsilon$ in the  following estimation
\begin{eqnarray}
f_{KYR} &=& \varepsilon\,m_K\sqrt{\Gamma_{\Delta(1940)\,\rightarrow\,K\Sigma}\,\frac{12\,\pi}{|\vec p_K\,|^3} \left(\frac{m_Y}{E_Y + m_Y}\right)}
\nonumber\\
&=& \varepsilon\,0.28\,,
\end{eqnarray}
where we have used $\Gamma_{\Delta(1940)\rightarrow K\Sigma} = 2.3\;{\rm MeV}$ as $\Gamma_{\Delta(1940)\rightarrow K\Sigma} = 2.0 - 2.5\;{\rm MeV}$ leads to a good fit to experimental data. It is found in our calculation that $\varepsilon = -1$, the relative sign of $f_{KYR}$\,, results in a better description of experimental data than $\varepsilon = 1$.
Moreover, we will use the hadronic form factor, $F_{\Delta(1940)}$, for $K^0\Sigma^+$ photoproduction with $\Delta(1940)$ resonance as,
\begin{eqnarray}
F_{\Delta(1940)} = \frac{\Lambda_\Delta^4}{(s - m_\Delta^2)^2 + \Lambda_\Delta^4}\,,
\end{eqnarray}
where $\Lambda_\Delta$ is the cut-off momentum of $\Delta(1940)$\,. In this work, the cut-off $\Lambda_\Delta = 500$ MeV is used. We will incorporate the $\Delta(1940)$ resonance amplitude with all parameters estimated here to the box-diagrams and the modified K-MAID in the next section.

\section{Numerical results}
In this section, we calculate the differential cross-section for the $\gamma p\,\rightarrow\,K^0\Sigma^+$ photoproduction process by including the contributions of the box-diagrams, the modified K-MAID and the $\Delta(1940)$ resonance and compare the results with the experimental data \cite{Ewald:2011gw}. In the original convention and notation in Ref. \cite{Chew:1957tf}, the differential cross-section in c.m. frame is defined by \cite{Chew:1957tf,Thom:1966rm}
\begin{eqnarray}
\frac{d\,\sigma}{d\,\Omega} = \frac{\widetilde{q}}{\widetilde{k}}\,\big|\,\mathcal{F}\,\big|^2\,,
\end{eqnarray}
with
\begin{widetext}
\begin{eqnarray}
\big|\mathcal{F}\big|^2&=& {\rm Re}\,\Big\{ \big|\mathcal{F}_1\big|^2 + \big|\mathcal{F}_2\big|^2
- 2\,\cos\theta\,\mathcal{F}_1^*\,\mathcal{F}_2 + \sin^2\theta\,\Big( \textstyle{\frac12}\,\big|\mathcal{F}_3\big|^2
+ \textstyle{\frac12}\,\big|\mathcal{F}_4\big|^2 + \mathcal{F}_1^*\,\mathcal{F}_4 + \mathcal{F}_2^*\,\mathcal{F}_3
+ \cos\theta\,\mathcal{F}_3^*\,\mathcal{F}_4\Big)  \Big\}.
\end{eqnarray}
\begin{figure}
\begin{tabular}{c}
\includegraphics[width=12cm,height=8cm,angle=0]{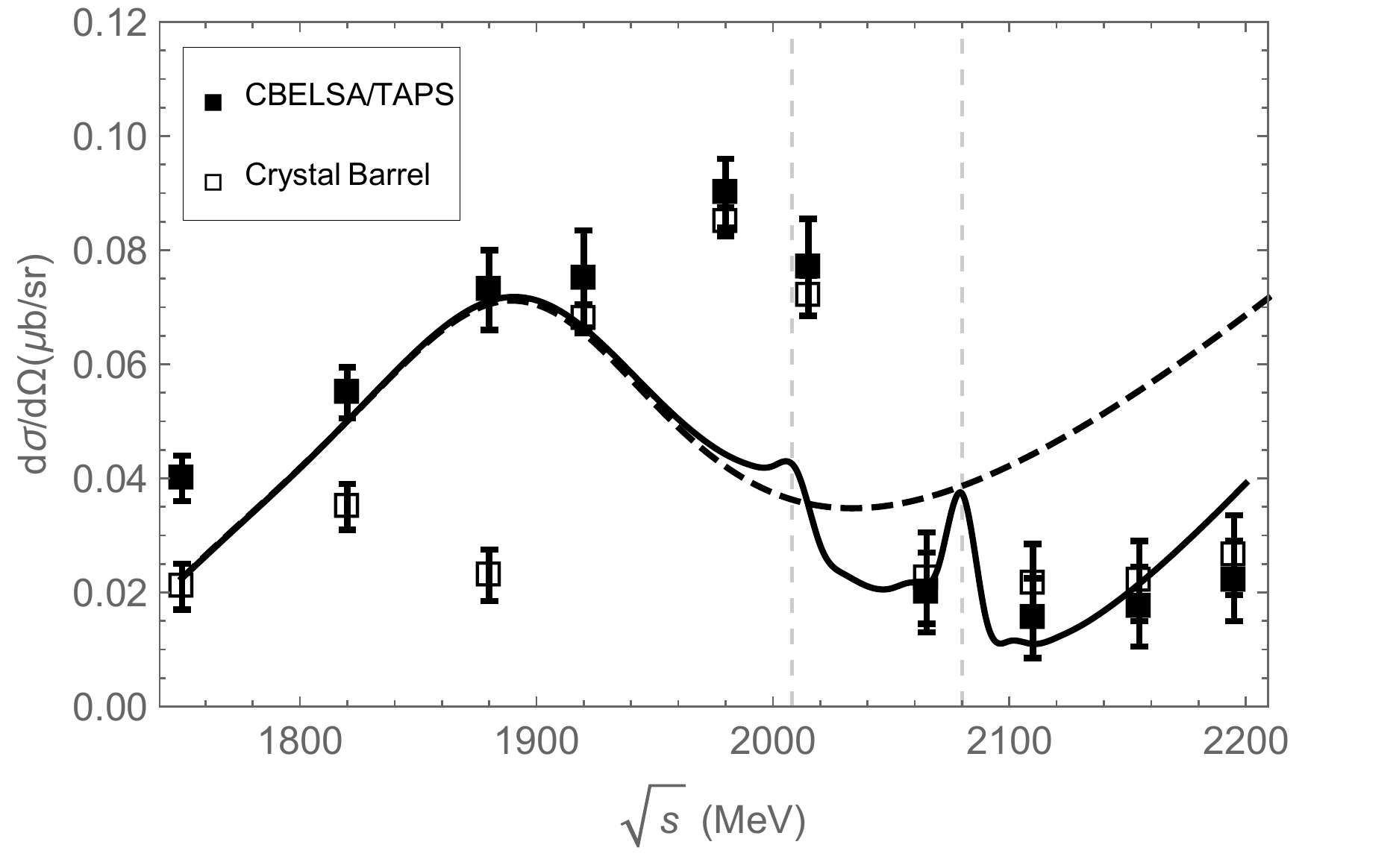}
\\
\includegraphics[width=12cm,height=8cm,angle=0]{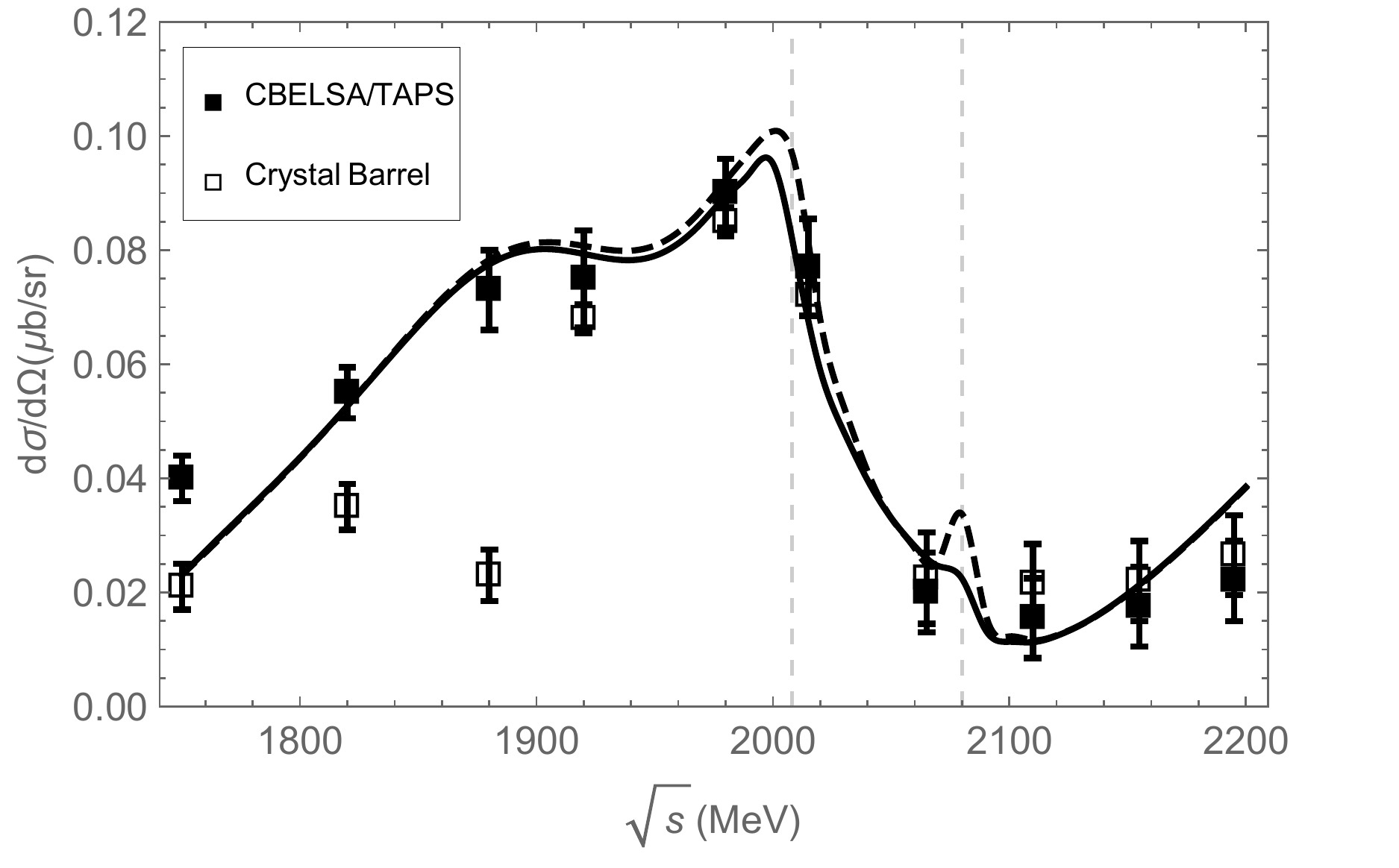}
\end{tabular}
\caption{Theoretical results of the differential cross section   with respect to CBELSA/TAPS (solid square) and Crystal Barrel \cite{Castelijns:2007qt} (open square). The upper panel shows the result from modified K-MAID (dashed line) with $G1_{\rm S_{31}}(1900) = 0.8\,,~ G1_{\rm P_{31}}(1910) = 0.6\,,~ G2_{\rm S_{31}}(1900) = 0.8\,,~ G2_{\rm P_{31}}(1910) = 0.6$ and the cobmined contributions (solid line) of the modified K-MAID model and the imaginary part of the box-diagrams with the best fitted hadronic phases, $\theta_1 = 37.74^\circ$\,, $\theta_2 = 47.14^\circ$\,, $\theta_3 = 47.82^\circ$. In the lower panel, the dash line is the result contributed by the modified K-MAID, the imaginary part of the box-diagrams, and the $\Delta(1940)$ resonance with the hadronic phases  $\theta_1 = 37.74^\circ$\,, $\theta_2 = 47.14^\circ$\,, $\theta_3 = 47.82^\circ$\, while the solid line is the result of the modified K-MAID, the full amplitudes (real and imaginary parts) of the box-diagrams, and the inclusion of the $\Delta(1940)$ resonance.}
\label{diff-cross-plots}
\end{figure}
%%%%%%%%%%%%%%%%%%%%%%%%%%%%%%%%%%%%%%%%%%%%%%%%%%%%%%%%%%%%%%%%%%%%%%%%%%%%%%%%%%%%%%%%%%%%%%%%%%%%%%%%%%%

%%%%%%%%%%%%%%%%%%%%%%%%%%%%%%%%%%%%%%%%%%%%%%%%%%%%%%%%%%%%%%%%%%%%%%%%%%%%%%%%%%%%%%%%%%%%%%%%%%%%%%%%%%%%%
%\begin{figure}
%\begin{tabular}{ccc}
%\includegraphics[width=5cm,height=4cm,angle=0]{plot1823.eps}
%&\includegraphics[width=5cm,height=4cm,angle=0]{plot1874.eps}
%&\includegraphics[width=5cm,height=4cm,angle=0]{plot1923.eps}
%\\
%\includegraphics[width=5cm,height=4cm,angle=0]{plot1971.eps}
%&\includegraphics[width=5cm,height=4cm,angle=0]{plot2018.eps}
%&\includegraphics[width=5cm,height=4cm,angle=0]{plot2064.eps}
%\\
%\includegraphics[width=5cm,height=4cm,angle=0]{plot2109.eps}
%&\includegraphics[width=5cm,height=4cm,angle=0]{plot2153.eps}
%&\includegraphics[width=5cm,height=4cm,angle=0]{plot2197.eps}
%\end{tabular}
%\caption{Differential cross section for the $\gamma\,p\,\rightarrow\,K^0\,\Sigma^+$
%reaction as a function of the kaon center-of-mass angle\,.}
%\label{diff-angle}
%\end{figure}
%%%%%%%%%%%%%%%%%%%%%%%%%%%%%%%%%%%%%%%%%%%%%%%%%%%%%%%%%%%%%%%%%%%%%%%%%%%%%%%%%%%%%%%%%%%%%%%%%%%%%%%%%%%%
The CGLN amplitudes $\mathcal{F}_i$ for the $\gamma p\,\rightarrow\,K^0\Sigma^+$ photoproduction process are given by
\begin{eqnarray}
\mathcal{F}_i = \alpha\,\mathcal{F}_i^{\rm KMAID} + \beta\,\mathcal{F}_i^{\Delta(1940)} + \delta\left(e^{i\,\theta_1}\,\mathcal{F}_i^{\rm box\,(1)} + e^{i\,\theta_2}\,\mathcal{F}_i^{\rm box\,(2)} + e^{i\,\theta_3}\,\mathcal{F}_i^{\rm box\,(3)}\right),
\end{eqnarray}
\end{widetext}
where $\mathcal{F}_i^{\rm KMAID}$, $\mathcal{F}_i^{\Delta(1940)}$ and $\mathcal{F}_i^{\rm box\,(1,2,3)}$ are the CGLN amplitudes for the modified K-MAID model, the $\Delta(1940)$ resonance and the imaginary part of the box diagrams 1, 2 and 3 respectively. We have introduced as free parameters the hadronic phases, $\theta_{1,2,3}$ to the box-diagram amplitudes. The parameters
are determined by fitting the contributions of the box diagrams
to experimental data,
\begin{eqnarray}
\theta_1 = 37.74^\circ\,,\qquad\quad \theta_2 = 47.14^\circ\,,\qquad\quad \theta_3 = 47.82^\circ\,.
\end{eqnarray}
In addition, the parameters $\alpha,\,\beta,\,\delta = 0,\,1$ are employed as switches to see the contributions of the K-MAID model, $\Delta(1940)$ resonance and box diagrams to the total amplitude respectively.

The numerical result of the imaginary part of the box-diagrams and K-MAID amplitudes are shown in the upper panel of Fig. \ref{diff-cross-plots} by setting $\alpha,\delta = 1$ and $\beta=0$\,. The downfall of the differential cross-section is contributed by the on-shell intermediate $K^*\Lambda$ and $K^*\Sigma$ states in their mass threshold region. This implies that there are re-scattering effects between the vector kaons and the hyperons in the intermediate states of the processes as depicted in Fig. 2. It is found from Fig. \ref{diff-cross-plots}, however, that the box-diagram and K-MAID amplitudes can not describe the data from 1920 MeV to the $K^*\Lambda$ threshold.

Shown in the lower panel of Fig. \ref{diff-cross-plots} are the results including the $\Delta(1940)$ resonance contribution, where the dash line is the result contributed by the modified K-MAID, the imaginary part of the box-diagrams, and the $\Delta(1940)$ resonance with the hadronic phases  $\theta_1 = 37.74^\circ$\,, $\theta_2 = 47.14^\circ$\,, $\theta_3 = 47.82^\circ$\, while the solid line is the result contributed together by the modified K-MAID, the full amplitudes (real and imaginary parts) of the box-diagrams, and the $\Delta(1940)$ resonance. One can see that the inclusion of the $\Delta(1940)$ resonance contribution largely raises the differential cross-section, leading to an excellent agreement with the anomaly of the $K^0\Sigma^+$ photoproduction data.

The bump structure may merely stem from the interference between the amplitudes as the inclusion of the real part of the box-diagrams contributions results in the disappearance of the structure.
%We further calculate the differential cross-section as a function of the kaon angle in the c.m. frame. The results are compared with the data and depicted in Fig. \ref{diff-angle}. With all contribution of the amplitudes, our results show that the modified K-MAID gives good agreement with the data at the forward angle. The $\Delta(1940)$ resonance starts its contribution at 1971 MeV and rises the different cross-section below the $K^*\Lambda$ threshold. The on-shell box-diagrams begin the contributions at $K^*\Lambda$ and the differential cross-section falls down beyond the threshold. By fitting the hadronic phases with the data, we obtain a good result at forward angle region. However, we note that the theoretical results in this work are better agreement with the data of the forward angles than the the backward angles. The reason of this phenomena is probably that we fit the free parameters (hadronic phases) in our model with the CBELSA/TAPS data of the
%$\gamma p\,\to\,K^0\Sigma^+$ differential cross-section at the forward angle of the kaon in c.m. frame, $\theta_K^{\rm c.m.}= 30^\circ$. To improve the backward angle results, one might take the $p$-wave contribution, other degrees of freedom of the reactions and etc. into account which are beyond the scope of this work.

\section{Conclusions}
%According to the speculation of the data in \cite{Ewald:2011gw}, the anomaly in the sudden drop of the differential cross-section in the $\gamma p\,\to\,K^0\Sigma^+$ process at $K^*\Lambda$ and $K^*\Sigma$ thresholds is probably occurred from the $K^*Y$ states.
In this work, we have calculated the contributions of the box-diagrams, the modified K-MAID model, and the $\Delta(1940)$ resonance to the differential cross-section of the photoproduction reaction $\gamma p\,\to\,K^0\Sigma^+$. The imaginary part of the amplitudes of the box-diagrams is obtained by using the Cutkosky rules while the real part is calculated by using the dispersion relations from the imaginary part of the box-diagram amplitudes.
The hadronic phases of the on-shell amplitudes of the box diagrams are introduced as free parameters determined by fitting the theoretical results with the CBELSA/TAPS data.

It is found that the modified K-MAID model provides a good description of the $K^0\Sigma^+$ photoproduction up to 1920 MeV,
and that the inclusion of the $\Delta(1940)$ resonance which contributes via the process $\gamma p\;\rightarrow\;\Delta(1940)\;\rightarrow\,K^0\Sigma^+$ is necessary to describe the anomaly of the CBELSA/TAPS data.

The work reveals that the theoretical results are sensitive to the property of the $\Delta(1940)$ resonance. Therefore, one may suggest that this two-star resonance may be further investigated in experiments via the
$\gamma p\,\to\,K^0\Sigma^+$ reaction.
 %However, there is bump like structure at the $K^*\Sigma$ threshold. The real part of the on-shell box-diagram amplitudes is used to get rid the bump (solid line) in the lower panel of Fig. \ref{diff-cross-plots}. In addition, %the differential cross-section in terms of the kaon angles in the c.m. frame is also presented. T
%In addition, the results also behave a nearly flat differential cross-section beyond the $K^*\Sigma$ threshold at the forward angles.

\acknowledgments
This work is supported by Suranaree University of Technology and the Office of the Higher Education Commission under NRU project of Thailand.
KX and YY acknowledge support from SUT under Grant
No. SUT-PhD/13/2554. XL is supported by Young Science Foundation from the Education Department of Liaoning Province, China (Project No. LQ2019009). DS is supported by Thailand research fund (TRF) under contract No. TRG6180014.

%%%%%%%%%%%%%%%%%%%%%%%%%%%%%%%%%%%%%%%%%%%%%%%%%%%%%%%%%%%%%%%%%%%%%%%%%%%%%%%%%%%%%%%%%%%%%%%%%%%%%%%%%%%%%%%%%%%%%%%%%%%%%%%%%%%%%%%%%%%%%%%%%%%%%

\appendix*
\section{Extraction of the CGLN amplitudes for $\gamma p \rightarrow K^0\Sigma^+$ box-diagrams}
We give in the section the detailed calculation of the on-shell box-diagrams in the $K^0\Sigma^+$ photoproduction process. As mentioned earlier, the CGLN amplitudes are invariant amplitude of the photoproduction in the c.m. frame, and can be conveniently incorporated with other sophisticated photoproduction models.
The CGLN amplitudes for the box-diagrams shown in Eqs. (\ref{box1}, \ref{box2}, \ref{box3}) will be extracted in the appendix.
%\subsection{Invariant amplitude}
We start from the most general Lorentz covariant transition matrix element for the photo-production process, which takes the generic form \cite{Berends:1967vi}
\begin{eqnarray}
\mathcal{M} & = & \bar u(\bar p)\,
\sum_{j=1}^6 B_{j}(s,t,u) \,\epsilon_\mu\,N_{j}^\mu\, u(p)~ ,
\label{invar-B}
\end{eqnarray}
where $B_j$ are scalar functions and $N_j^\mu$ are four-vector quantities given by
\begin{eqnarray}
\mathcal{N}_{1}^{\mu} &=& \gamma_{5}\,\gamma^{\mu}\,k\!\!\!/ \,,\quad
\mathcal{N}_{2}^{\mu} = 2\gamma_{5}\,P^{\mu} \,,\quad
\mathcal{N}_{3}^{\mu} = 2\gamma_{5}\,q^{\mu} \,,
\nonumber\\
\mathcal{N}_{4}^{\mu} &=& \gamma_{5}\,\gamma^{\mu} \,,\quad
\mathcal{N}_{5}^{\mu} = \gamma_{5}\,k\!\!\!/\,P^{\mu} \,,\quad
\mathcal{N}_{6}^{\mu} = \gamma_{5}\,k\!\!\!/\,q^{\mu} \,.
\end{eqnarray}
One notes that the internal momentum, $l^\mu$ always appears in the on-shell amplitudes (\ref{box1}, \ref{box2}, \ref{box3}) and it will appear on the basis $N_i^\mu$\,. To factorize the $l^\mu$ momentum out from the basis $N_i^\mu$, we decompose $l^\mu$ in the following way by using Lorentz invariant principle as,
\begin{eqnarray}
l^\mu &=& a\,k^\mu + b\,q^\mu + c\,p^\mu \,,
\end{eqnarray}
where $a,~b,$\, and $c$ are scalar functions, derived as
\begin{widetext}
\begin{eqnarray}
a &=& -\frac{p^2 q^2 k\cdot l-p^2 k\cdot q l\cdot q-q^2 k\cdot p l\cdot p-k\cdot l (p\cdot q)^2+k\cdot q l\cdot p p\cdot q+k\cdot p l\cdot q p\cdot q}{p^2 (k\cdot q)^2+q^2 (k\cdot p)^2-2 k\cdot p k\cdot q p\cdot q},
\nonumber\\
b &=& -\frac{-p^2 k\cdot l k\cdot q+k\cdot p k\cdot q l\cdot p+(k\cdot p)^2 (-(l\cdot q))+k\cdot l k\cdot p p\cdot q}{p^2 (k\cdot q)^2+q^2 (k\cdot p)^2-2 k\cdot p k\cdot q p\cdot q},
\nonumber\\
c &=& -\frac{-q^2 k\cdot l k\cdot p+(k\cdot q)^2 (-(l\cdot p))+k\cdot p k\cdot q l\cdot q+k\cdot l k\cdot q p\cdot q}{p^2 (k\cdot q)^2+q^2 (k\cdot p)^2-2 k\cdot p k\cdot q p\cdot q}\,.
\end{eqnarray}
In addition, the incoming, $p$, and outgoing, $\bar p$, momenta of baryons are also replaced by using the following substitution
\begin{eqnarray}
p^\mu &=& \frac12\,\big( 2\,P^\mu + q^\mu - k^\mu \big)\,,
\nonumber\\
\bar p^\mu &=& \frac12\,\big( 2\,P^\mu - q^\mu + k^\mu \big)\,,
\end{eqnarray}
where we have defined $P = \frac12\,(p + \bar p)$\,.

By using the above definitions for $l^\mu,~p^\mu$ and $\bar p^\mu$\,, and substituting to Eqs. (\ref{box1},\ref{box2},\ref{box3}), one obtains the decomposition of the amplitudes (\ref{box1},\ref{box2},\ref{box3}) in terms of the scalar functions $B_i$ defined in Eq. (\ref{invar-B}) by using the FeynCalc package \cite{Mertig:1990an,Shtabovenko:2016sxi} in Mathematica. The scalar functions $B_i$ are given by
\begin{eqnarray}
B_1^{(i)} &=& \int\rho_i\,\Big\{-m_p m_V^4 -m_{\Sigma } m_V^4 -b m_p^3 m_V^2 -c m_p^3 m_V^2 -b m_p m_{\Sigma }^2 m_V^2 +b m_Y m_{\Sigma }^2 m_V^2 -b m_K^2 m_p m_V^2 -c m_K^2 m_p m_V^2
\nonumber\\
&& -\,m_K^2 m_p m_V^2 +b k\cdot q m_p m_V^2 +c k\cdot q m_p m_V^2 +2 k\cdot q m_p m_V^2 +2 a k\cdot \bar{p} m_p m_V^2 +2 b k\cdot \bar{p} m_p m_V^2 +2 l\cdot p m_p m_V^2
\nonumber\\
&& +\,b l\cdot q m_p m_V^2 +c l\cdot q m_p m_V^2 +2 l\cdot q m_p m_V^2 +2 b p\cdot \bar{p} m_p m_V^2 +m_K^2 m_Y m_V^2 +b m_p^2 m_Y m_V^2
+c m_p^2 m_Y m_V^2
\nonumber\\
&& -\,a k\cdot q m_Y m_V^2 -b k\cdot q m_Y m_V^2 -2 k\cdot q m_Y m_V^2 -2 a k\cdot \bar{p} m_Y m_V^2 -2 b k\cdot \bar{p} m_Y m_V^2 -l\cdot q m_Y m_V^2 -2 b p\cdot \bar{p} m_Y m_V^2
\nonumber\\
&& -\,b m_K^2 m_{\Sigma } m_V^2 -c m_p^2 m_{\Sigma } m_V^2 -a k\cdot q m_{\Sigma } m_V^2 +2 l\cdot p m_{\Sigma } m_V^2 +b l\cdot q m_{\Sigma } m_V^2 +l\cdot q m_{\Sigma } m_V^2 +c m_p m_Y m_{\Sigma } m_V^2
\nonumber\\
&& +\,b k\cdot q m_p^3 +c k\cdot q m_p^3 +b l\cdot q m_p^3 +c l\cdot q m_p^3 +b k\cdot q m_p m_{\Sigma }^2 +b l\cdot q m_p m_{\Sigma }^2
-b k\cdot q m_Y m_{\Sigma }^2 -b l\cdot q m_Y m_{\Sigma }^2
\nonumber\\
&& +\,2 b l\cdot p m_K^2 m_p +2 c l\cdot p m_K^2 m_p -2 a k\cdot q k\cdot \bar{p} m_p -2 b k\cdot q k\cdot \bar{p} m_p +2 a k\cdot q l\cdot p m_p -2 c k\cdot q l\cdot p m_p -2 a k\cdot q l\cdot q m_p
\nonumber\\
&& -\,2 b k\cdot q l\cdot q m_p -2 k\cdot q l\cdot q m_p -2 a k\cdot \bar{p} l\cdot q m_p -2 b k\cdot \bar{p} l\cdot q m_p -2 b l\cdot p l\cdot q m_p -2 c l\cdot p l\cdot q m_p -2 l\cdot p l\cdot q m_p
\nonumber\\
&& -\,2 b k\cdot q p\cdot \bar{p} m_p - 2 c k\cdot q p\cdot \bar{p} m_p -2 b l\cdot q p\cdot \bar{p} m_p -b k\cdot q m_p^2 m_Y -c k\cdot q m_p^2 m_Y -b l\cdot q m_p^2 m_Y -c l\cdot q m_p^2 m_Y
\nonumber\\
&& +\,2 a k\cdot q k\cdot \bar{p} m_Y +2 b k\cdot q k\cdot \bar{p} m_Y +2 a k\cdot q l\cdot q m_Y +2 b k\cdot q l\cdot q m_Y +2 k\cdot q l\cdot q m_Y +2 a k\cdot \bar{p} l\cdot q m_Y
\nonumber\\
&& +\,2 b k\cdot \bar{p} l\cdot q m_Y + 2 b k\cdot q p\cdot \bar{p} m_Y +2 c k\cdot q p\cdot \bar{p} m_Y +2 b l\cdot q p\cdot \bar{p} m_Y +2 b l\cdot p m_K^2 m_{\Sigma } -c k\cdot q m_p^2 m_{\Sigma }
\nonumber\\
&& +\,c l\cdot q m_p^2 m_{\Sigma } +2 a k\cdot q l\cdot p m_{\Sigma } -2 b l\cdot p l\cdot q m_{\Sigma } -2 l\cdot p l\cdot q m_{\Sigma } +c k\cdot q m_p m_Y m_{\Sigma } -c l\cdot q m_p m_Y m_{\Sigma }
\nonumber\\
&& -\,2 (k\cdot l)^2 \left((a-c) m_p+a m_{\Sigma }\right)
+2 k\cdot p \,\big(-b m_p m_K^2 -c m_p m_K^2 -b m_{\Sigma } m_K^2 -a m_p m_V^2 -b m_p m_V^2 -m_p m_V^2
\nonumber\\
&& +\,a m_V^2 m_Y+b m_V^2 m_Y-m_V^2 m_{\Sigma }+l\cdot q \left((a+2 b+c+1) m_p-(a+b) m_Y+(b+1) m_{\Sigma }\right) +k\cdot q \big((b+c) m_p
\nonumber\\
&& -\,(a+b) m_Y - a m_{\Sigma }\big)\big)+k\cdot l \big( b m_p^3 +c m_p^3-b m_Y m_p^2 -c m_Y m_p^2 +c m_{\Sigma } m_p^2 +a m_K^2 m_p +3 b m_K^2 m_p +2 c m_K^2 m_p
\nonumber\\
&& +\,m_K^2 m_p + a m_V^2 m_p -c m_V^2 m_p +m_V^2 m_p +b m_{\Sigma }^2 m_p -2 b k\cdot \bar{p} m_p -2 a l\cdot p m_p +2 c l\cdot p m_p
-2 b l\cdot q m_p -2 c l\cdot q m_p
\nonumber\\
&& -\,4 l\cdot q m_p -2 b p\cdot \bar{p} m_p -c m_Y m_{\Sigma } m_p -b m_Y m_{\Sigma }^2 -a m_K^2 m_Y -b m_K^2 m_Y -m_K^2 m_Y +m_V^2 m_Y
+2 b k\cdot \bar{p} m_Y
\nonumber\\
&& +\,2 l\cdot q m_Y+2 b p\cdot \bar{p} m_Y+2 b m_K^2 m_{\Sigma }+a m_V^2 m_{\Sigma }+2 m_V^2 m_{\Sigma }-2 a l\cdot p m_{\Sigma }-2 b l\cdot q m_{\Sigma }-2 l\cdot q m_{\Sigma }
\nonumber\\
&& +\,2 k\cdot q \left((a-c) m_p+a m_{\Sigma }\right)+2 k\cdot p \left((a+b) m_p-(b+c) m_Y+a m_{\Sigma }\right)\big)\Big\}\,,
\nonumber\\
%%%%%%%%%%%%%%%%%%%%%%%%%%%%%%%%%%%%%%%%%%%%%%%%%%%%%%%%%%%%%%%%%%%%%%%%%%%%%%%%%%%%%%%%%%%%%%%%
%%%%%%%%%%%%%%%%%%%%%%%%%%%%%%%%%%%%%%%%%%%%%%%%%%%%%%%%%%%%%%%%%%%%%%%%%%%%%%%%%%%%%%%%%%%%%%%%
B_2^{(i)} &=& \int\rho_i\,\frac{1}{2}\, \Big\{2 k\cdot l \big(2 b c m_{\Sigma } k\cdot \bar{p}+2 b c m_p k\cdot \bar{p}+c k\cdot q \left((b+c-2) m_p+b m_{\Sigma }\right)-2 c k\cdot p \left((b+c) m_p+(b-1) m_{\Sigma }\right)
\nonumber\\
&& -\,b m_K^2 m_Y +2 c m_p l\cdot p-c m_p m_V^2-m_p m_V^2+m_V^2 m_Y\big)+2 b c m_K^2 m_Y k\cdot \bar{p}-2 b c m_K^2 m_p k\cdot \bar{p}-2 b c m_{\Sigma } k\cdot q k\cdot \bar{p}
\nonumber\\
&& -\,2 b c m_Y k\cdot q k\cdot \bar{p} -2 b c m_Y p\cdot \bar{p} k\cdot q+2 b c m_p p\cdot \bar{p} k\cdot q-4 c^2 m_p l\cdot p k\cdot \bar{p}+2 c^2 m_p m_V^2 k\cdot \bar{p}- 2 c m_V^2 m_Y k\cdot \bar{p}
\nonumber\\
&&+2 c m_p m_V^2 k\cdot \bar{p}+2 c k\cdot p k\cdot q \left((c-a) m_p+(a-c+1) m_Y+(c-1) m_{\Sigma }\right)-2 b c m_{\Sigma } k\cdot q l\cdot p-2 b c m_p k\cdot q l\cdot p
\nonumber\\
&& -\,b c m_p m_{\Sigma }^2 k\cdot q+b c m_p m_V^2 k\cdot q+b c m_p^2 m_Y k\cdot q-b c m_p^3 k\cdot q+b c m_{\Sigma } m_V^2 k\cdot q+b c m_{\Sigma }^2 m_Y k\cdot q-4 b m_p k\cdot q l\cdot q
\nonumber\\
&& +\,4 b m_Y k\cdot q l\cdot q+2 b m_p m_V^2 k\cdot q-2 b m_V^2 m_Y k\cdot q-2 c^2 m_p k\cdot q l\cdot p-c^2 m_p^2 m_{\Sigma } k\cdot q+c^2 m_p m_V^2 k\cdot q
\nonumber\\
&& +\,c^2 m_p m_{\Sigma } m_Y k\cdot q+c^2 m_p^2 m_Y k\cdot q-c^2 m_p^3 k\cdot q-4 c^2 m_{\Sigma } (k\cdot p)^2+2 c m_{\Sigma } k\cdot q l\cdot p-2 c m_p k\cdot q l\cdot p+4 c m_p (k\cdot l)^2
\nonumber\\
&& +\,2 c m_p m_V^2 k\cdot q-c m_{\Sigma } m_V^2 k\cdot q-c m_V^2 m_Y k\cdot q+2 m_p m_V^2 k\cdot q-2 m_V^2 m_Y k\cdot q +b m_K^2 m_p\Big\}\,,
\nonumber\\
%%%%%%%%%%%%%%%%%%%%%%%%%%%%%%%%%%%%%%%%%%%%%%%%%%%%%%%%%%%%%%%%%%%%%%%%%%%%%%%%%%%%%%%%%%%%%%%%
%%%%%%%%%%%%%%%%%%%%%%%%%%%%%%%%%%%%%%%%%%%%%%%%%%%%%%%%%%%%%%%%%%%%%%%%%%%%%%%%%%%%%%%%%%%%%%%%
B_3^{(i)} &=& \int\rho_i\,\frac{1}{4} \Big\{-2 b^2 k\cdot q m_p^3-c^2 k\cdot q m_p^3-3 b c k\cdot q m_p^3+2 b^2 k\cdot q m_Y m_p^2+c^2 k\cdot q m_Y m_p^2+3 b c k\cdot q m_Y m_p^2
\nonumber\\
&& -\,2 b c k\cdot q m_{\Sigma } m_p^2-4 b^2 k\cdot \bar{p} m_K^2 m_p-2 b c k\cdot \bar{p} m_K^2 m_p+2 b^2 k\cdot q m_V^2 m_p+c^2 k\cdot q m_V^2 m_p-2 b k\cdot q m_V^2 m_p
\nonumber\\
&& +\,3 b c k\cdot q m_V^2 m_p-2 c k\cdot q m_V^2 m_p-2 k\cdot q m_V^2 m_p+2 c^2 k\cdot \bar{p} m_V^2 m_p+4 b c k\cdot \bar{p} m_V^2 m_p-2 c k\cdot \bar{p} m_V^2 m_p
\nonumber\\
&& -\,2 b^2 k\cdot q m_{\Sigma }^2 m_p-b c k\cdot q m_{\Sigma }^2 m_p-4 b^2 k\cdot q l\cdot p m_p-2 c^2 k\cdot q l\cdot p m_p+4 b k\cdot q l\cdot p m_p-6 b c k\cdot q l\cdot p m_p
\nonumber\\
&& +\,6 c k\cdot q l\cdot p m_p-4 c^2 k\cdot \bar{p} l\cdot p m_p-8 b c k\cdot \bar{p} l\cdot p m_p+8 c k\cdot \bar{p} l\cdot p m_p+4 b k\cdot q l\cdot q m_p+8 b k\cdot \bar{p} l\cdot q m_p
\nonumber\\
&& +\,4 b^2 k\cdot q p\cdot \bar{p} m_p+2 b c k\cdot q p\cdot \bar{p} m_p+c^2 k\cdot q m_Y m_{\Sigma } m_p+2 b c k\cdot q m_Y m_{\Sigma } m_p+2 b^2 k\cdot q m_Y m_{\Sigma }^2+b c k\cdot q m_Y m_{\Sigma }^2
\nonumber\\
&& +\,4 b^2 k\cdot \bar{p} m_K^2 m_Y+2 b c k\cdot \bar{p} m_K^2 m_Y-c k\cdot q m_V^2 m_Y+2 k\cdot q m_V^2 m_Y-2 c k\cdot \bar{p} m_V^2 m_Y+4 k\cdot \bar{p} m_V^2 m_Y
\nonumber\\
&& -\,4 b^2 k\cdot q k\cdot \bar{p} m_Y -2 b c k\cdot q k\cdot \bar{p} m_Y-4 b k\cdot q l\cdot q m_Y-8 b k\cdot \bar{p} l\cdot q m_Y-4 b^2 k\cdot q p\cdot \bar{p} m_Y-2 b c k\cdot q p\cdot \bar{p} m_Y
\nonumber\\
&& -\,4 c (2 b+c-2) (k\cdot p)^2 m_{\Sigma } +2 b^2 k\cdot q m_V^2 m_{\Sigma }-2 b k\cdot q m_V^2 m_{\Sigma }+b c k\cdot q m_V^2 m_{\Sigma }-c k\cdot q m_V^2 m_{\Sigma }-4 b^2 k\cdot q k\cdot \bar{p} m_{\Sigma }
\nonumber\\
&& -\,2 b c k\cdot q k\cdot \bar{p} m_{\Sigma }-4 b^2 k\cdot q l\cdot p m_{\Sigma }+4 b k\cdot q l\cdot p m_{\Sigma }-2 b c k\cdot q l\cdot p m_{\Sigma }+2 c k\cdot q l\cdot p m_{\Sigma } +4 b (k\cdot l)^2 \left(m_p+m_{\Sigma }\right)
\nonumber\\
&& +\,2 k\cdot p k\cdot q \left(-(a-c) (2 b+c) m_p+(a-c+1) (2 b+c) m_Y+(2 (c+1) b+(c-1) c) m_{\Sigma }\right)+2 k\cdot l \big(b m_p^3+c m_p^3
\nonumber\\
&& -b m_Y m_p^2-c m_Y m_p^2+c m_{\Sigma } m_p^2-b m_K^2 m_p-b m_V^2 m_p+m_V^2 m_p+b m_{\Sigma }^2 m_p+4 b^2 k\cdot \bar{p} m_p-4 b k\cdot \bar{p} m_p
\nonumber\\
&& +\,2 b c k\cdot \bar{p} m_p +2 b l\cdot p m_p -2 l\cdot p m_p-2 b p\cdot \bar{p} m_p-c m_Y m_{\Sigma } m_p-b m_Y m_{\Sigma }^2+b m_K^2 m_Y+2 b k\cdot \bar{p} m_Y+2 b p\cdot \bar{p} m_Y
\nonumber\\
&& -\,b m_V^2 m_{\Sigma }+m_V^2 m_{\Sigma } +4 b^2 k\cdot \bar{p} m_{\Sigma }-2 b k\cdot \bar{p} m_{\Sigma }+2 b c k\cdot \bar{p} m_{\Sigma }+2 b l\cdot p m_{\Sigma }-2 l\cdot p m_{\Sigma } -4 k\cdot \bar{p} m_V^2 m_p
\nonumber\\
&& +\,k\cdot q \big(\left(2 b^2+(3 c-4) b+(c-2) c\right) m_p+(2 b+c-4) b m_{\Sigma }\big)+2 k\cdot p \big(\left(-2 b^2-3 c b+2 b-c^2+a+c\right) m_p
\nonumber\\
&& +\,(-a+c-1) m_Y-\left(2 b^2+(c-2) b+1\right) m_{\Sigma }\big)\big) -c^2 k\cdot q m_{\Sigma } m_p^2 \Big\}\,,
\nonumber\\
%%%%%%%%%%%%%%%%%%%%%%%%%%%%%%%%%%%%%%%%%%%%%%%%%%%%%%%%%%%%%%%%%%%%%%%%%%%%%%%%%%%%%%%%%%%%%%%%%%%%%%%
%%%%%%%%%%%%%%%%%%%%%%%%%%%%%%%%%%%%%%%%%%%%%%%%%%%%%%%%%%%%%%%%%%%%%%%%%%%%%%%%%%%%%%%%%%%%%%%%%%%%%%%
B_4^{(i)} &=& \int\rho_i\,\Big\{-(k\cdot q) m_V^4-2 k\cdot \bar{p} m_V^4-2 b k\cdot \bar{p} m_K^2 m_V^2-2 b k\cdot q m_p^2 m_V^2-2 c k\cdot q m_p^2 m_V^2-k\cdot q m_p^2 m_V^2
\nonumber\\
&& -\,b k\cdot q m_{\Sigma }^2 m_V^2+4 b k\cdot q k\cdot \bar{p} m_V^2+2 k\cdot q l\cdot p m_V^2+4 k\cdot \bar{p} l\cdot p m_V^2+2 k\cdot q l\cdot q m_V^2+2 b k\cdot \bar{p} l\cdot q m_V^2
\nonumber\\
&& +\,2 k\cdot \bar{p} l\cdot q m_V^2+2 b k\cdot q p\cdot \bar{p} m_V^2+b k\cdot q m_p m_Y m_V^2+c k\cdot q m_p m_Y m_V^2+k\cdot q m_p m_Y m_V^2+2 c k\cdot \bar{p} m_p m_Y m_V^2
\nonumber\\
&& +\,b k\cdot q m_p m_{\Sigma } m_V^2+c k\cdot q m_p m_{\Sigma } m_V^2+k\cdot q m_p m_{\Sigma } m_V^2-b k\cdot q m_Y m_{\Sigma } m_V^2-k\cdot q m_Y m_{\Sigma } m_V^2
\nonumber\\
&& +\,4 b k\cdot \bar{p} l\cdot p m_K^2+2 b k\cdot q l\cdot p m_p^2+2 c k\cdot q l\cdot p m_p^2+2 b k\cdot q l\cdot q m_p^2+2 c k\cdot q l\cdot q m_p^2+2 c k\cdot \bar{p} l\cdot q m_p^2
\nonumber\\
&& +\,2 b k\cdot q l\cdot p m_{\Sigma }^2-4 b k\cdot q k\cdot \bar{p} l\cdot p-4 k\cdot q l\cdot p l\cdot q-4 b k\cdot \bar{p} l\cdot p l\cdot q-4 k\cdot \bar{p} l\cdot p l\cdot q -2 c k\cdot \bar{p} m_p^2 m_V^2
\nonumber\\
&& +\,4 (k\cdot p)^2 \left(-b m_K^2-c m_K^2-m_V^2+(b+c) k\cdot q+(b+c+1) l\cdot q\right)-4 b k\cdot q l\cdot p p\cdot \bar{p}-2 (k\cdot l)^2 \big(b m_p^2+c m_p^2
\nonumber\\
&& -\,c m_{\Sigma } m_p+b m_{\Sigma }^2 +2 (a+b) k\cdot p-2 b k\cdot \bar{p}-2 b p\cdot \bar{p}\big)+k\cdot l \big(4 (a+b) (k\cdot p)^2+2 \big(b m_K^2+c m_K^2+m_K^2+b m_p^2
\nonumber\\
&& +\,c m_p^2+a m_V^2+b m_V^2+2 m_V^2+b m_{\Sigma }^2+2 (a+b) k\cdot q -c m_K^2 m_p^2-2 b l\cdot p m_p^2
\nonumber\\
&& -\,2 b k\cdot \bar{p}-2 a l\cdot p-2 b l\cdot p-4 l\cdot q-2 b p\cdot \bar{p}-c m_Y m_{\Sigma }\big) k\cdot p+2 b k\cdot \bar{p} m_K^2+2 l\cdot p m_K^2-b m_K^2 m_p^2
\nonumber\\
&& -\,2 c l\cdot p m_p^2-m_K^2 m_V^2+b m_p^2 m_V^2+c m_p^2 m_V^2+m_p^2 m_V^2-2 b k\cdot \bar{p} m_V^2-2 b p\cdot \bar{p} m_V^2+b m_V^2 m_{\Sigma }^2-2 b l\cdot p m_{\Sigma }^2
\nonumber\\
&& +\,4 b k\cdot \bar{p} l\cdot p-4 b k\cdot \bar{p} l\cdot q+4 b l\cdot p p\cdot \bar{p}+2 k\cdot q \left(b m_p^2+c m_p^2-c m_{\Sigma } m_p+b m_{\Sigma }^2-2 b k\cdot \bar{p}-2 b p\cdot \bar{p}\right)
\nonumber\\
&& -\,m_p m_V^2 m_Y+b m_K^2 m_p m_Y+c m_K^2 m_p m_Y-c m_p m_V^2 m_{\Sigma }-m_p m_V^2 m_{\Sigma }+b m_K^2 m_p m_{\Sigma }+2 c l\cdot p m_p m_{\Sigma }
\nonumber\\
&& -\,b m_K^2 m_Y m_{\Sigma }+m_V^2 m_Y m_{\Sigma }\big) -2 b k\cdot q l\cdot q m_p m_Y-2 c k\cdot q l\cdot q m_p m_Y-2 c k\cdot \bar{p} l\cdot q m_p m_Y
\nonumber\\
&& -\,2 c k\cdot q l\cdot p m_p m_{\Sigma }-2 b k\cdot q l\cdot q m_p m_{\Sigma } +2 b k\cdot q l\cdot q m_Y m_{\Sigma }
\nonumber\\
&& -\,2 k\cdot p k\cdot q \big(-b m_p^2-c m_p^2+a m_V^2+2 b m_V^2+c m_V^2-m_V^2-b m_{\Sigma }^2+2 (a+b) k\cdot \bar{p}-2 (a+b) l\cdot p+2 l\cdot q
\nonumber\\
&& +\,2 b p\cdot \bar{p}+2 c p\cdot \bar{p}-c m_Y m_{\Sigma }\big)\Big\}\,,
\nonumber\\
%%%%%%%%%%%%%%%%%%%%%%%%%%%%%%%%%%%%%%%%%%%%%%%%%%%%%%%%%%%%%%%%%%%%%%%%%%%%%%%%%%%%%%%%%%%%%%%%%%%%%%%
%%%%%%%%%%%%%%%%%%%%%%%%%%%%%%%%%%%%%%%%%%%%%%%%%%%%%%%%%%%%%%%%%%%%%%%%%%%%%%%%%%%%%%%%%%%%%%%%%%%%%%%
B_5^{(i)} &=& \int\rho_i\,\Big\{2 m_V^4+2 b m_K^2 m_V^2+c m_K^2 m_V^2-c^2 m_p^2 m_V^2-b c m_p^2 m_V^2+c m_p^2 m_V^2-b c m_{\Sigma }^2 m_V^2+2 a k\cdot q m_V^2
\nonumber\\
&& +\,b c k\cdot q m_V^2 -c k\cdot q m_V^2+2 a c k\cdot \bar{p} m_V^2+2 b c k\cdot \bar{p} m_V^2-4 l\cdot p m_V^2-2 b l\cdot q m_V^2-2 l\cdot q m_V^2+2 b c p\cdot \bar{p} m_V^2
\nonumber\\
&& -\,c m_p m_Y m_V^2+c^2 m_p m_{\Sigma } m_V^2+c m_p m_{\Sigma } m_V^2-c m_Y m_{\Sigma } m_V^2+4 a (k\cdot l)^2-4 c (b+c) (k\cdot p)^2-4 b l\cdot p m_K^2
\nonumber\\
&& -\,2 c l\cdot p m_K^2+c^2 m_K^2 m_p^2+b c m_K^2 m_p^2-2 c^2 k\cdot q m_p^2+a c k\cdot q m_p^2-b c k\cdot q m_p^2+c k\cdot q m_p^2+2 c^2 l\cdot p m_p^2
\nonumber\\
&& +\,2 b c l\cdot p m_p^2-2 c l\cdot q m_p^2+2 b c l\cdot p m_{\Sigma }^2-2 b c k\cdot q k\cdot \bar{p}-4 a k\cdot q l\cdot p-2 a c k\cdot q l\cdot p-2 b c k\cdot q l\cdot p
\nonumber\\
&& +\,2 c k\cdot q l\cdot p-4 a c k\cdot \bar{p} l\cdot p-4 b c k\cdot \bar{p} l\cdot p+4 b l\cdot p l\cdot q+4 l\cdot p l\cdot q
+a c k\cdot q m_V^2 -2 k\cdot p \big(m_p^2 c^2-m_p m_{\Sigma } c^2
\nonumber\\
&& -\,m_K^2 c+b m_p^2 c+a m_V^2 c+b m_V^2 c+b m_{\Sigma }^2 c-2 b k\cdot \bar{p} c-2 a l\cdot p c-2 b l\cdot p c-2 b p\cdot \bar{p} c-2 b m_K^2
\nonumber\\
&& -\,2 m_V^2+(a (c-2)-(c-1) c) k\cdot q+2 b l\cdot q+2 l\cdot q\big)+2 k\cdot l \big(m_p^2 c^2-m_p m_{\Sigma } c^2-m_K^2 c+b m_p^2 c+b m_{\Sigma }^2 c
\nonumber\\
&& -\,2 b p\cdot \bar{p} c-m_p m_{\Sigma } c+m_Y m_{\Sigma } c-2 b m_K^2-a m_V^2-2 m_V^2-2 (a-c) k\cdot p+((c-2) a+b c) k\cdot q+2 a l\cdot p
\nonumber\\
&& +\,2 b l\cdot q+2 l\cdot q\big) -4 b c l\cdot p p\cdot \bar{p}-c^2 m_K^2 m_p m_Y-b c m_K^2 m_p m_Y+2 c^2 k\cdot q m_p m_Y-a c k\cdot q m_p m_Y
\nonumber\\
&& +\,2 c l\cdot q m_p m_Y-b c m_K^2 m_p m_{\Sigma }-a c k\cdot q m_p m_{\Sigma }+b c k\cdot q m_p m_{\Sigma }+c k\cdot q m_p m_{\Sigma }-2 c^2 l\cdot p m_p m_{\Sigma }
\nonumber\\
&& +\,b c m_K^2 m_Y m_{\Sigma }+a c k\cdot q m_Y m_{\Sigma }-b c k\cdot q m_Y m_{\Sigma }-c k\cdot q m_Y m_{\Sigma } +b c k\cdot q m_p m_Y-c k\cdot q m_p m_Y \Big\}\,,
\nonumber\\
%%%%%%%%%%%%%%%%%%%%%%%%%%%%%%%%%%%%%%%%%%%%%%%%%%%%%%%%%%%%%%%%%%%%%%%%%%%%%%%%%%%%%%%%%%%%%%%%%%%%%%%
%%%%%%%%%%%%%%%%%%%%%%%%%%%%%%%%%%%%%%%%%%%%%%%%%%%%%%%%%%%%%%%%%%%%%%%%%%%%%%%%%%%%%%%%%%%%%%%%%%%%%%%
B_6^{(i)} &=& \int\rho_i\,\frac{1}{2} \Big\{ 2 m_K^2 m_p^2 b^2-2 k\cdot q m_p^2 b^2+4 l\cdot p m_p^2 b^2-2 m_p^2 m_V^2 b^2+2 k\cdot q m_V^2 b^2+4 k\cdot \bar{p} m_V^2 b^2+4 p\cdot \bar{p} m_V^2 b^2
\nonumber\\
&&+\,4 l\cdot p m_{\Sigma }^2 b^2-4 k\cdot q k\cdot \bar{p} b^2-4 k\cdot q l\cdot p b^2-8 k\cdot \bar{p} l\cdot p b^2-8 l\cdot p p\cdot \bar{p} b^2-2 m_K^2 m_p m_Y b^2+2 k\cdot q m_p m_Y b^2
\nonumber\\
&& -\,2 m_K^2 m_p m_{\Sigma } b^2+2 k\cdot q m_p m_{\Sigma } b^2+2 m_K^2 m_Y m_{\Sigma } b^2-2 k\cdot q m_Y m_{\Sigma } b^2+4 (k\cdot l)^2 b+3 c m_K^2 m_p^2 b+2 a k\cdot q m_p^2 b
\nonumber\\
&& -\,5 c k\cdot q m_p^2 b+2 k\cdot q m_p^2 b+6 c l\cdot p m_p^2 b-4 l\cdot p m_p^2 b-4 l\cdot q m_p^2 b-3 c m_p^2 m_V^2 b+2 m_p^2 m_V^2 b+2 a k\cdot q m_V^2 b+c k\cdot q m_V^2 b
\nonumber\\
&& -\,2 k\cdot q m_V^2 b+4 a k\cdot \bar{p} m_V^2 b+2 c k\cdot \bar{p} m_V^2 b-4 k\cdot \bar{p} m_V^2 b+2 l\cdot q m_V^2 b+2 c p\cdot \bar{p} m_V^2 b-4 p\cdot \bar{p} m_V^2 b -c m_V^2 m_{\Sigma }^2 b
\nonumber\\
&&+\,2 m_V^2 m_{\Sigma }^2 b+2 c l\cdot p m_{\Sigma }^2 b-4 l\cdot p m_{\Sigma }^2 b-2 c k\cdot q k\cdot \bar{p} b-4 a k\cdot q l\cdot p b-2 c k\cdot q l\cdot p b +4 k\cdot q l\cdot p b-8 a k\cdot \bar{p} l\cdot p b
\nonumber\\
&&-\,4 c k\cdot \bar{p} l\cdot p b+8 k\cdot \bar{p} l\cdot p b-4 l\cdot p l\cdot q b-4 c l\cdot p p\cdot \bar{p} b+8 l\cdot p p\cdot \bar{p} b -3 c m_K^2 m_p m_Y b-2 a k\cdot q m_p m_Y b
\nonumber\\
&&+\,5 c k\cdot q m_p m_Y b-2 k\cdot q m_p m_Y b+4 l\cdot q m_p m_Y b+2 c m_p m_V^2 m_{\Sigma } b-c m_K^2 m_p m_{\Sigma } b-2 a k\cdot q m_p m_{\Sigma } b
\nonumber\\
&& +\,c k\cdot q m_p m_{\Sigma } b-2 k\cdot q m_p m_{\Sigma } b-4 c l\cdot p m_p m_{\Sigma } b+4 l\cdot q m_p m_{\Sigma } b+c m_K^2 m_Y m_{\Sigma } b+2 a k\cdot q m_Y m_{\Sigma } b
\nonumber\\
&&-\,c k\cdot q m_Y m_{\Sigma } b+2 k\cdot q m_Y m_{\Sigma } b-4 l\cdot q m_Y m_{\Sigma } b-4 \left(2 b^2+(3 c+2) b+c^2+2 a\right) (k\cdot p)^2 -2 m_V^2 m_{\Sigma }^2 b^2
\nonumber\\
&& -\,2 c l\cdot p m_K^2+c^2 m_K^2 m_p^2-2 c^2 k\cdot q m_p^2+a c k\cdot q m_p^2+c k\cdot q m_p^2+2 c^2 l\cdot p m_p^2-4 c l\cdot p m_p^2-2 c l\cdot q m_p^2+c m_K^2 m_V^2
\nonumber\\
&& -\,c^2 m_p^2 m_V^2+c m_p^2 m_V^2+2 m_p^2 m_V^2-2 a k\cdot q m_V^2+a c k\cdot q m_V^2-c k\cdot q m_V^2-4 a k\cdot \bar{p} m_V^2+2 a c k\cdot \bar{p} m_V^2-2 l\cdot q m_V^2
\nonumber\\
&& +\,4 a k\cdot q l\cdot p-2 a c k\cdot q l\cdot p+2 c k\cdot q l\cdot p+8 a k\cdot \bar{p} l\cdot p-4 a c k\cdot \bar{p} l\cdot p+4 l\cdot p l\cdot q+c m_p m_V^2 m_Y-2 m_p m_V^2 m_Y
\nonumber\\
&& -\,c^2 m_K^2 m_p m_Y+2 c^2 k\cdot q m_p m_Y-a c k\cdot q m_p m_Y-c k\cdot q m_p m_Y+2 c l\cdot q m_p m_Y+c^2 m_p m_V^2 m_{\Sigma }-c m_p m_V^2 m_{\Sigma }
\nonumber\\
&& -\,2 m_p m_V^2 m_{\Sigma } -a c k\cdot q m_p m_{\Sigma }+c k\cdot q m_p m_{\Sigma }-2 c^2 l\cdot p m_p m_{\Sigma }+4 c l\cdot p m_p m_{\Sigma }-c m_V^2 m_Y m_{\Sigma }+2 m_V^2 m_Y m_{\Sigma }
\nonumber\\
&& +\,a c k\cdot q m_Y m_{\Sigma } -c k\cdot q m_Y m_{\Sigma }-2 k\cdot l \big(-2 m_p^2 b^2-2 m_{\Sigma }^2 b^2+4 p\cdot \bar{p} b^2-3 c m_p^2 b+m_p^2 b+m_V^2 b-c m_{\Sigma }^2 b+2 m_{\Sigma }^2 b
\nonumber\\
&& -\,2 k\cdot \bar{p} b -2 l\cdot p b+2 l\cdot q b+2 c p\cdot \bar{p} b-4 p\cdot \bar{p} b+m_p m_Y b+2 c m_p m_{\Sigma } b+m_p m_{\Sigma } b-m_Y m_{\Sigma } b+c m_K^2-c^2 m_p^2+a m_p^2
\nonumber\\
&& +\,m_p^2-m_V^2 +2 k\cdot p-((2 b+c-2) a+(2 b+c-4) b) k\cdot q+2 l\cdot p-2 l\cdot q-a m_p m_Y+2 c m_p m_Y-m_p m_Y
\nonumber\\
&& +\,c^2 m_p m_{\Sigma }-a m_p m_{\Sigma }-c m_p m_{\Sigma }-m_p m_{\Sigma }+a m_Y m_{\Sigma } -c m_Y m_{\Sigma }+m_Y m_{\Sigma }\big)+2 k\cdot p \big(-2 m_p^2 b^2-2 m_V^2 b^2
\nonumber\\
&& -\,2 m_{\Sigma }^2 b^2+4 l\cdot p b^2+4 p\cdot \bar{p} b^2-3 c m_p^2 b-2 m_p^2 b-2 a m_V^2 b-c m_V^2 b +2 m_V^2 b-c m_{\Sigma }^2 b-2 m_{\Sigma }^2 b +4 a l\cdot p b+2 c l\cdot p b
\nonumber\\
&&-\,4 l\cdot p b+2 l\cdot q b+2 c p\cdot \bar{p} b+4 p\cdot \bar{p} b+2 c m_p m_{\Sigma } b+c m_K^2-c^2 m_p^2 -2 c m_p^2+2 a m_V^2-a c m_V^2
\nonumber\\
&&+\,(-(2 b+c-2) a+2 (c+1) b+(c-1) c) k\cdot q+2 (2 a+(2 b+c+2) b) k\cdot \bar{p}-4 a l\cdot p+2 a c l\cdot p+2 l\cdot q
\nonumber\\
&& +\,4 c p\cdot \bar{p}+c^2 m_p m_{\Sigma }-2 c m_p m_{\Sigma }\big)\Big\} \,,
\end{eqnarray}
\end{widetext}
where we have defined the notation $\int\rho_i$ as
\begin{eqnarray}
\int\rho_i = \frac{\widetilde{l}}{32\,\pi^2\,\sqrt{s}}\,G_i \int d\,\Omega\;\frac{F_{K}^2(k-l)}{(k-l)^2 - m_{\bar K}^2}
\,\frac{F_\pi^2(l - q) }{(l - q)^2 -m_\pi^2}\,.
\end{eqnarray}
In addition, the following identities are applied for factorizing the $B_i^{(j)}$ functions,
\begin{eqnarray}
l^\mu &=& l_{\nu}\,g^{\mu\nu} = \frac12\,(\gamma^\mu\,l\!\!\!/ + l\!\!\!/\,\gamma^\mu) \,,
\\
\epsilon^{\mu\nu\alpha\beta}\,\gamma_{\beta} &=& i\,(g^{\mu\nu}\,\gamma^\alpha - g^{\mu\alpha}\,\gamma^\nu + g^{\nu\alpha}\,\gamma^\mu - \gamma^\mu\,\gamma^\nu\,\gamma^\alpha)\,\gamma_5 \,.
\nonumber
\end{eqnarray}
The photoproduction amplitude $\mathcal{M}$ is conventionally decomposed into the gauge- and
Lorentz-invariant matrices $M_i$ where we have used the convention in Refs. \cite{Berends:1967vi}. The $\mathcal{M}$ can be written as
\begin{eqnarray}
\mathcal{M} & = & \bar u(\bar p)\,
\sum_{j=1}^4 A_{j}(s,t,u) \,M_{j}\, u(p)~ ,
\end{eqnarray}
where $s$ and $t$ are the usual Mandelstam variables, defined by
\begin{eqnarray}
s = (k+p)^2 ,~~ t = (k-q)^2 ,~~ u ~=~ (k - \bar p)^2 ~,
\label{mandelstam}
\end{eqnarray}
The gauge and Lorentz invariant matrices $M_{j}$ are given by
\begin{eqnarray}
M_{1} &=& {\textstyle \frac{1}{2}}\, \gamma_{5}\, \big\{ \epsilon \!\!\!/\, k
\!\!\!/ - k \!\!\!/ \,\epsilon \!\!\!/ \big\}~ ,
\nonumber\\
M_{2} &=& 2\,\gamma_{5}\,\big\{ (q \cdot \epsilon)\,( P \cdot k) -
(q\cdot k)\,( P \cdot \epsilon) \big\}~ ,
\nonumber\\
M_{3} &=& \gamma_{5} \,\big\{ (q\cdot k)\, \epsilon \!\!\!/ - (q\cdot \epsilon)\, k
\!\!\!/ \big\} ~ ,
\\
M_{4} &=& 2\,\gamma_5\,\big\{ (P\cdot k)\,\epsilon\!\!\!/ - (P\cdot \epsilon)\,k\!\!\!/\big\} - (m_p + m_\Sigma)\,M_1\,,
\nonumber
\end{eqnarray}
where $P = \frac{1}{2}(p + \bar p)$, and $\epsilon_{\mu \nu \rho \sigma}$ is the four dimensional Levi-Civita tensor with $\epsilon_{0123} = +1$.

By imposing the gauge invariant condition, $k^\mu\epsilon_\mu^{(\gamma)}(k) =0$,  the $A_{i}$ are rewritten in terms of the scalar function $B_{i}$ via the following relations \cite{Berends:1967vi},
\begin{eqnarray}
A_{1} &=& B_{1}-\frac{1}{2}\,(m_{p}+m_{\Sigma})\,B_{5} \,,
\nonumber\\
A_{2} &=& \frac{2\,B_{2}}{M_{K}^{2}-t} \,,
\nonumber\\
A_{3} &=& -B_{6} \,,
\nonumber\\
A_{4} &=& -\frac{1}{2}B_{5} \,.
\end{eqnarray}
By using above relations, we obtain invariant amplitudes in terms of the Lorentz covariant and gauge invariant manners. However, this form of the Lorentz covariant fashion is not convenient for the non-relativistic calculation, e.g. photoproduction amplitudes form quark models. Then, we further proceed to extract the CGLN amplitudes which are defined by the following decomposition, \cite{Chew:1957tf}
\begin{widetext}
\begin{eqnarray}
\mathcal{F} = \mathcal{F}_1\,\vec\sigma\cdot\vec\varepsilon + \mathcal{F}_2\,i\,(\vec\sigma\cdot\hat q\,)\,(\vec\varepsilon\times\hat k\,)\cdot\,\vec\sigma + \mathcal{F}_3\,(\vec\sigma\cdot\hat k\,)\,(\hat q\cdot\vec\varepsilon\,) + \mathcal{F}_4\,(\vec\sigma\cdot\hat q\,)\,(\hat q\cdot\vec\varepsilon\,)\,,
\label{CGLN}
\end{eqnarray}
Therefore, the scalar functions, $\mathcal{F}_j$ of the CGLN amplitudes are given in terms of $A_j$ functions by \cite{Mart:2015jof}
\begin{eqnarray}
\mathcal{F}_{1,2} &=& \frac{1}{8\,\pi\,W}\,\sqrt{(E_p \mp m_p)\,(E_\Sigma \pm m_\Sigma)}\,\Big[ \pm\,(W \mp m_p)\,A_1 + \vec q\cdot \vec k\,(A_3 - A_4) + (W\mp m_p)\,(W\mp m_\Sigma)\,A_4 \Big]\,,
\\
\mathcal{F}_{3,4} &=& \frac{\widetilde{q}\,\widetilde{k}}{8\,\pi\,W}\,\sqrt{\frac{(E_\Sigma \pm m_\Sigma)}{(E_p \mp m_p)}}\,\Big[ \pm\,(s - m_p^2)\,A_2 + (W\mp m_p)\,(A_3 - A_4) \Big]\,.
\end{eqnarray}
\end{widetext}
The CGLN amplitudes of the on-shell box-diagrams are obtained in the analytical forms, and ready for numerical integrations as mentioned in Section III.

\bibliographystyle{apsrev4-1}

\bibliography{Kphoto-bib}

\end{document}